\pdfoutput=1
\documentclass[useAMS,usenatbib]{mn2e}

\usepackage{amssymb}
\usepackage{amsmath}
\usepackage[dvipdfmx]{graphicx}	
\usepackage{bmpsize}
\usepackage{float}
\usepackage{tabulary}
\usepackage{rotating}
\usepackage{caption}
\usepackage{cmap}
\usepackage[usenames, dvipsnames]{color}
\usepackage{lscape}
\usepackage{pdflscape}
\usepackage[titletoc,title]{appendix}
\usepackage{soul}
\usepackage{hyperref}

\bibpunct{(}{)}{;}{a}{,}{,}

\newcommand{\corrdre}[1]{{#1}}
\newcommand{\corr}[1]{{#1}}
\newcommand{\cdre}[1]{{#1}}

\definecolor{darkgreen}{rgb}{0.1, 0.6, 0.1}
\title[Dependence of Convective Boundary Mixing]{Dependence of Convective Boundary Mixing on Boundary Properties and Turbulence Strength}
\author[A. Cristini, R. Hirschi, C. Meakin, D. Arnett, C. Georgy, I. Walkington]{A. Cristini$^{a}$\thanks{E-mail: a.j.cristini@keele.ac.uk}, R. Hirschi$^{a,b}$, C. Meakin$^{c,d}$, D. Arnett$^c$, 
C. Georgy$^{e,a}$, I. Walkington$^{a}$\\\\
$^a$\textsf{Astrophysics Group, Keele University, Lennard-Jones Laboratories, Keele, ST5 5BG, UK}\\
$^b$\textsf{Kavli IPMU (WPI), The University of Tokyo, Kashiwa, Chiba 277-8583, Japan}\\
$^c$\textsf{Department of Astronomy, University of Arizona, Tucson, AZ 85721, USA}\\
$^d$\textsf{Karagozian \& Case, Inc., 700 N. Brand Blvd. Suite 700, Glendale, CA, 91203, USA}\\
$^e$\textsf{Geneva Observatory, University of Geneva, Ch. Maillettes 51, 1290 Versoix, Switzerland}}
\vspace{-1cm}
\begin{document}

\setlength{\abovedisplayskip}{5mm}
\setlength{\belowdisplayskip}{5mm}

\date{Accepted 2019 January 25. Received 2019 January 25; in original form 2017 September 18}

\pagerange{\pageref{firstpage}--\pageref{lastpage}} \pubyear{2017}

\maketitle

\label{firstpage}

\begin{abstract} 
Convective boundary mixing is one of the major uncertainties in stellar evolution. In order to study its dependence on boundary properties and turbulence strength in a controlled way, we computed a series of 3D hydrodynamical simulations of stellar convection during carbon burning with a varying boosting factor of the driving luminosity. Our 3D implicit large eddy simulations were computed with the \textsc{prompi} code. We performed a mean field analysis of the simulations  
within the Reynolds-averaged Navier-Stokes framework. Both the vertical RMS velocity within the convective region and the bulk Richardson number of the boundaries are found to scale with the driving luminosity as expected from theory: $v\propto L^{1/3}$ and  Ri$_{\textrm{B}}\propto L^{-2/3}$, respectively. The positions of the convective boundaries were estimated through the composition profiles across them, and the strength of convective boundary mixing was determined by analysing the boundaries within the framework of the entrainment law. We find that the entrainment is approximately inversely proportional to the bulk Richardson number, Ri$_{\textrm{B}}$ ($\propto \textrm{Ri}_{\textrm{B}}^{-\alpha}, \alpha \sim 0.75$). Although the entrainment law does not encompass all the processes occurring at boundaries, our results support the use of the entrainment law to describe convective boundary mixing in 1D models, at least for the advanced phases. The next steps and challenges ahead are also discussed.

\end{abstract}


\section{Introduction}
\label{intro}

One-dimensional (1D) stellar evolution models are currently the only computational tool that can be used to simulate the structure and evolution of a star from the zero-age main sequence (ZAMS) up to their final stage. Such models are invaluable due to their ability to provide estimates for stellar masses \citep{2014AJ....148...68B}, nucleosynthesis yields \citep{2016ApJS..225...24P}, progenitor structures \citep{2011ApJ...733...78A} and evolutionary ages \citep{2014A&A...566A...7N}. Due to their limited dimensionality, 1D stellar models must prescribe or approximate multi-dimensional phenomena. This often leads to a loss of predictability as such parameterisations usually contain free parameters which are tuned in order for the models to match observations (e.g. the mixing length theory; \citealt{1958ZA.....46..108B}). In order to model these highly complex, non-linear processes such as convection, three-dimensional (3D) hydrodynamic models can be computed which solve the equations of fluid motion (e.g. the Euler equations). Such simulations can test the underlying physical assumptions in the stellar models under semi-realistic astrophysical, macroscopic, but not necessarily microscopic conditions. These simulations resolve the dynamical time-scales associated with the fluid flow, whereas stellar models resolve the secular (thermal) time-scales.
Previous 3D hydrodynamic models of the carbon and oxygen burning shells in massive stars \citep{2007ApJ...667..448M,2013ApJ...769....1V,2017MNRAS.471..279C}, have revealed enlightening results: the dynamic properties of the convective flow; the balance between turbulent driving and \corr{kinetic energy} dissipation (viscosity); convective boundary mixing through turbulent entrainment\footnote{\citet{1973Sci...180.1356T} defines entrainment as the transport of fluid across an interface between two bodies of fluid by a shear-induced turbulent flux.}; propagation of g-mode waves throughout the stable regions; and some level of agreement with the entrainment law (Eq. \ref{entr_law}). Building upon our previous work producing the first 3D hydrodynamic models of the carbon shell \citep{2017MNRAS.471..279C}, we extend this by studying the effects of varying the driving luminosity within the same initial set-up, provided by the same stellar evolution model of a 15\,M$_\odot$ star.\\
\indent The structure of the paper is as follows. \corrdre{In Section 2, we briefly describe the set-up of the initial conditions for the hydrodynamical models. The results and their analyses are presented in Section 3. We conclude our findings in Section 4. 
The Appendices include more details about the computational tools used, various derivations and supplementary analyses using the RANS framework, with a particular emphasis on the turbulent kinetic energy.}

\begin{table*}
\begin{center}
\begin{tabulary}{\textwidth}{l || c c c c c c c c}
\hline \hline\\
& \textsf{eps1} & \textsf{eps33} & \textsf{eps100} & \textsf{eps333} & \textsf{eps1k} & \textsf{eps3k} & \textsf{eps10k} & \textsf{eps33k} \\\\
\hline\hline\\
\textsf{$\epsilon_{fac}$} & 1 & 33 & 100 & 333 & 1000 & 3000 & 1$\times10^4$ & 3.3$\times10^4$ \\\\
\textsf{$\tau_{sim}$} & 5000 & 5000 & 3600 & 1640 & 6000 & 3635 & 2000 & 1190 \\\\
\textsf{$v_{rms}$} & 6.80$\times10^5$ & 1.15$\times10^6$ & 1.59$\times10^6$ & 1.92$\times10^6$ & 4.56$\times10^6$ & 6.93$\times10^6$ & 1.03$\times10^7$ & 1.57$\times10^7$ \\\\
\textsf{$\tau_q$} & 3908 & 3900 & 2489 & 636 & 5003 & 2743 & 1498 & 592 \\\\
\textsf{$\tau_c$} & 3257 & 1789 & 1317 & 1025 & 465 & 316 & 219 & 153 \\\\
\textsf{$\tau_q/\tau_c$} & 1.20 & 2.18 & 1.89 & 0.62 & 10.76 & 8.68 & 6.84 & 3.87 \\\\
\textsf{$Ri_{B,u}$} & 1876 & 559 & 310 & 200 & 42 & 19 & 7 & 4 \\\\
\textsf{$Ri_{B,l}$} & 2.93$\times10^4$ & 8089 & 4203 & 3101 & 435 & 188 & 101 & 44 \\\\
\textsf{Ma} & 0.0028 & 0.0048 & 0.0070 & 0.0074 & 0.0209 & 0.0321 & 0.0481 & 0.0727 \\\\
\end{tabulary}
\caption[Summary of simulations properties for the \textsf{eps1} - \textsf{eps33k} models]{Summary of simulation properties. $\epsilon_{fac}$: nuclear energy generation rate boosting factor, $\tau_{\rm sim}$: simulated physical time, $v_{\rm rms}$: global RMS 
convective velocity, $\tau_q$: 
quasi-steady state time (total simulation time minus initial transient time), $\tau_c$: 
convective turnover time, $\textrm{Ri}_\textrm{B,u}$, $\textrm{Ri}_\textrm{B,l}$: bulk Richardson number for the upper and lower convective boundary regions, respectively, Ma: Mach number.}
\label{eps_tab}
\end{center}
\end{table*}

\begin{figure}
\includegraphics[width=0.5\textwidth]{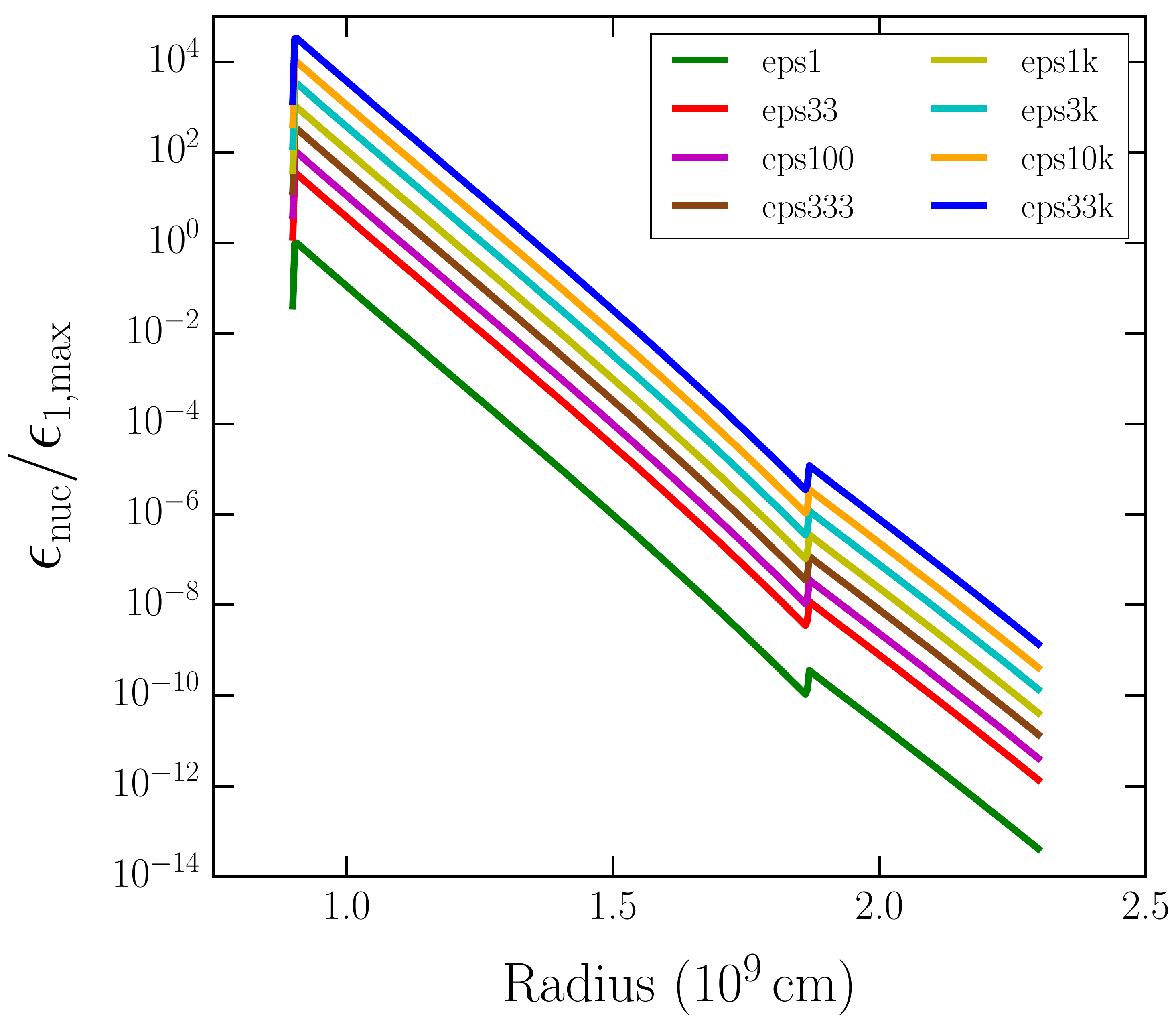}
\caption[Initial $\epsilon_{nuc}$ profiles of the \textsf{eps1} - \textsf{eps33k} models]{Radial profiles of the nuclear energy generation rate for each model. Each profile is normalised to the maximum energy generation rate in the nominal \textsf{eps1} model, to allow an easier comparison between the models.}
\label{eps_fac_pro}
\end{figure} 


\begin{figure}
\includegraphics[width=0.5\textwidth]{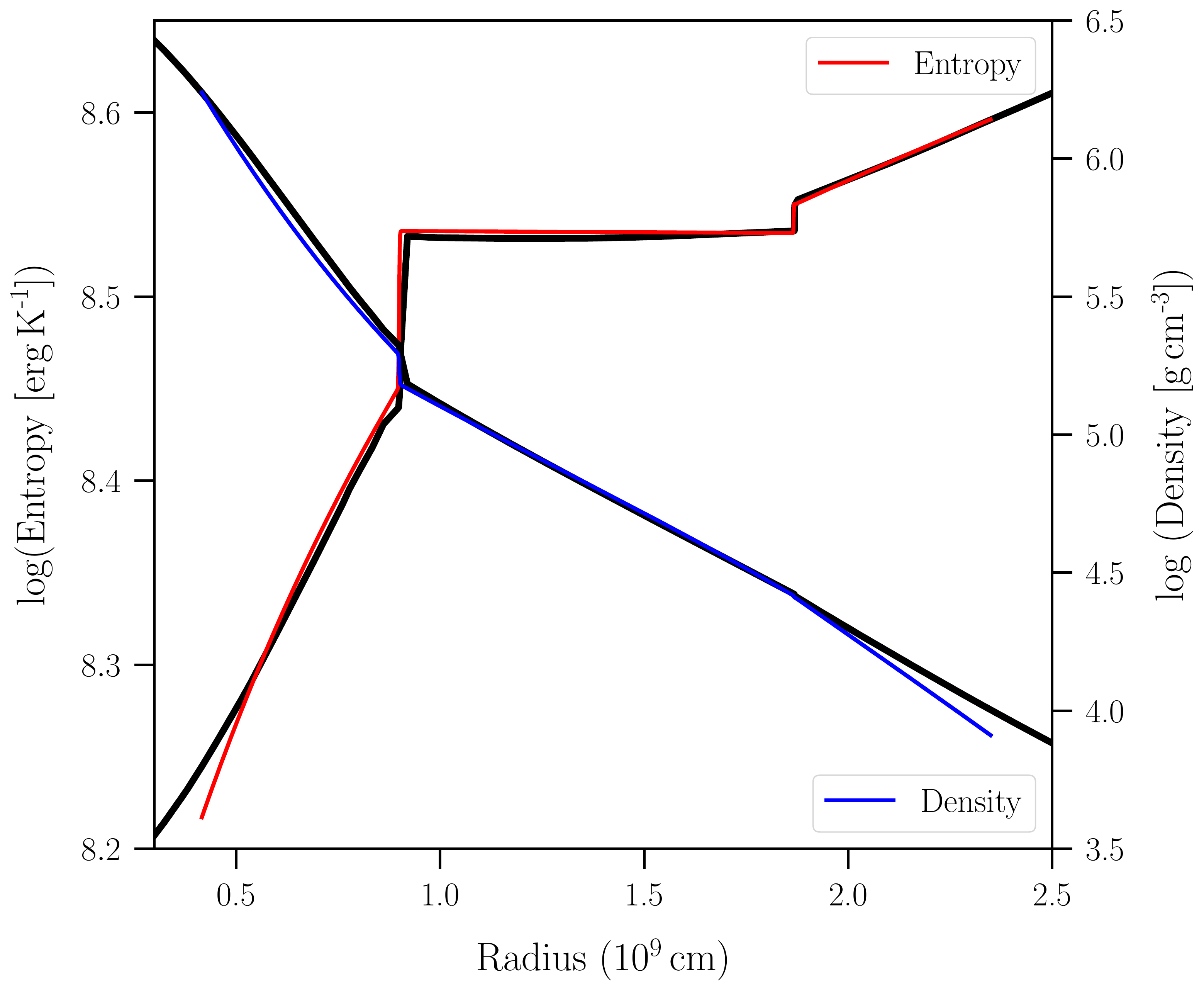}
\caption{Initial radial log density (blue) and log entropy (red) profiles. One dimensional stellar evolution 
profiles calculated using \textsc{genec} (black) are compared with the same profiles integrated and mapped onto the Eulerian Cartesian grid in \textsc{prompi} for the \textsf{eps1k} model.}
\label{eps_rho_entropy_log}
\end{figure} 

\section{Initial physical model}

\corrdre{The computational tools used to calculate the 
stellar model providing the initial conditions
and hydrodynamic simulations are described in Appendix \ref{comp_tools}.}

\corrdre{As in our previous study \citep[][C17 hereinafter]{2017MNRAS.471..279C}, we follow the ``box-in-star'' approach \citep[e.g.][]{2015ApJ...809...30A}, use a Cartesian coordinate system and a plane-parallel geometry. Details of the stellar model initial conditions can be found in \citet{2016PhyS...91c4006C} \cdre{and in Appendix \ref{comp_tools}}. The computational domain represents a convective region of thickness, $t$, bounded either side by radiative regions of thickness, $t/2$. Initially, the computational domain and convective region span $5.7$ and $2.6$ pressure scale heights, respectively. By including parts of the surrounding radiative regions, in general (the \textsf{eps33k} model is an exception) this ensures that over several convective turnovers the convective boundaries will not interact with the vertical domain boundaries. The aspect ratio of the convective zone is 2:1 (width:height), and therefore the plane-parallel approximation is not ideal. As in C17 this choice was made to ease the difficult Courant time scale condition at the inner boundary of the grid, allowing longer run-times as well as better resolution near convective boundaries. Direct comparison with oxygen burning simulations \citep[e.g.][]{2007ApJ...667..448M}, which use a spherical grid, suggest that no significant error results.}

\corrdre{The computational domain uses reflective, stress-free boundary conditions in the vertical direction and periodic boundary conditions in the two 
horizontal directions.  Although the material in the radiative regions is stable against convection it has oscillatory g-mode motions excited by 
the adjacent convection zone. \citet{2006ApJ...637L..53M} showed that such waves in their 2D model of the carbon and oxygen shell are well described by the linearised non-radial wave equation.} 
\corrdre{If left to propagate freely and due to the reflective boundary conditions, standing waves will develop which will affect the hydrostatic balance. Therefore, in order to mimic the propagation of these waves out of the domain, a damping region is employed which extends radially between a radius of $0.6\times10^9$\,cm 
and the lower domain boundary at $0.42\times10^9$\,cm. The damping region covers the full horizontal extent of the computational domain between these radii. Within this region all velocity components are reduced by a common damping factor, $f$, resulting in damped velocities over the damping region, $\boldsymbol{v_{d}}=f\boldsymbol{v}$. The damping factor is defined as}

\begin{equation}
f=\left(1+\delta t\, \omega f_{d}\right)^{-1},
\end{equation}

\corrdre{where $\delta t$ is the time step of the simulation, $\omega=0.01$ is the damping frequency and is a free parameter chosen to correspond to a small fraction of the convective turnover. $f_{d}=0.5\;(\textrm{cos}\left(\pi r/r_0\right)+1)$, where $r$ is the radial position in the vertical direction and $r_0 = 0.6\times10^9$\,cm is the edge of the damping region in the vertical direction. Using this damping function, $f_d =0$ at $r=r_0$, where the damping region starts. This ensures a smooth transition between the non-damped and damped regions.}

\begin{figure*}
\includegraphics[width=0.75\textwidth]{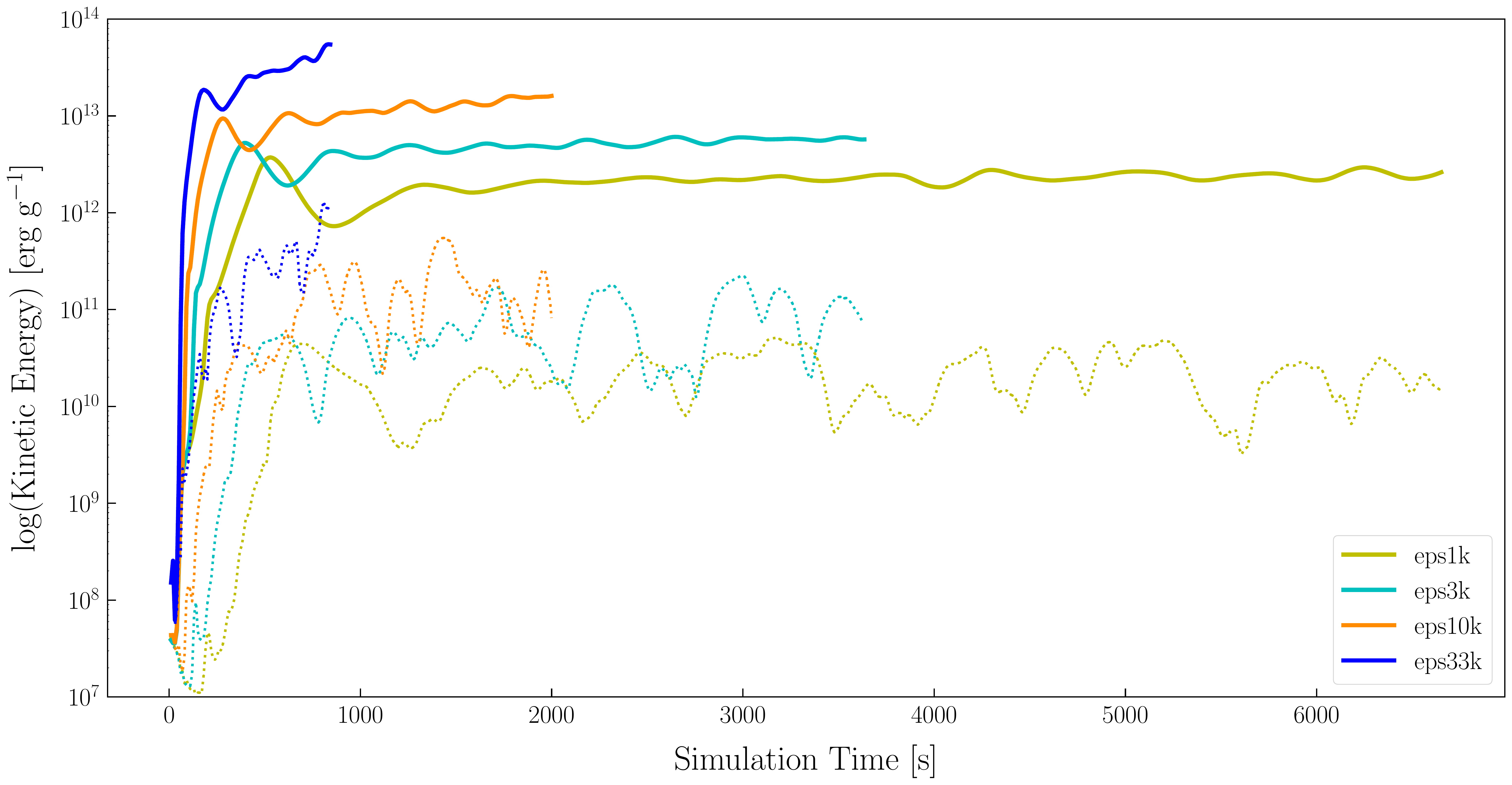}
\caption{Specific kinetic energy (thick solid) integrated over the computational domain for each timestep of the models \textsf{eps1k}, \textsf{eps3k}, \textsf{eps10k} and \textsf{eps33k}. The initial transient is characterised by a sharp rise to a local maximum, followed by a shallow decrease. The roughly horizontal parts of each profile that follow correspond to the quasi-steady turbulent phases of each model. Apart from the \textsf{eps33k} model, the end of each profile indicates the end of each simulation. The thin dotted lines show the integrated kinetic energy over the computational domain minus the integrated turbulent kinetic energy over the computational domain. These lines therefore represent the kinetic energy associated with energy transfer that is not a result of turbulent convection, i.e. kinetic energy due to waves propagating throughout the computational domain.}
\label{ke_and_ke-tke}
\end{figure*} 

\corrdre{The energy generation rate is calculated using the same prescription as described by eq. 6 in C17, with the addition of a constant boosting factor, $\epsilon_{fac}$. This factor was set to 1000 in C17. In this study, $\epsilon_{fac}$ varies between $1$ (denoted model \textsf{eps1}) and $3.3\times10^4$ (denoted model \textsf{eps33k}). Except for the energy generation rate, all of the models in this study are identical in set-up to the \textsf{hrez} model of C17. The radial profile of the nuclear energy generation rate is plotted for all models in Fig. \ref{eps_fac_pro}, and are normalised to the maximum energy generation rate of the nominal luminosity model. The energy generation rate due to neutrino losses is unchanged for each model, as over such short (dynamical) time-scales it is not expected that changes in the energy generation rate due to neutrino losses will have any considerable effect on the structure of the shell.}

\begin{figure*}
\centering
\includegraphics[width=0.75\textwidth]{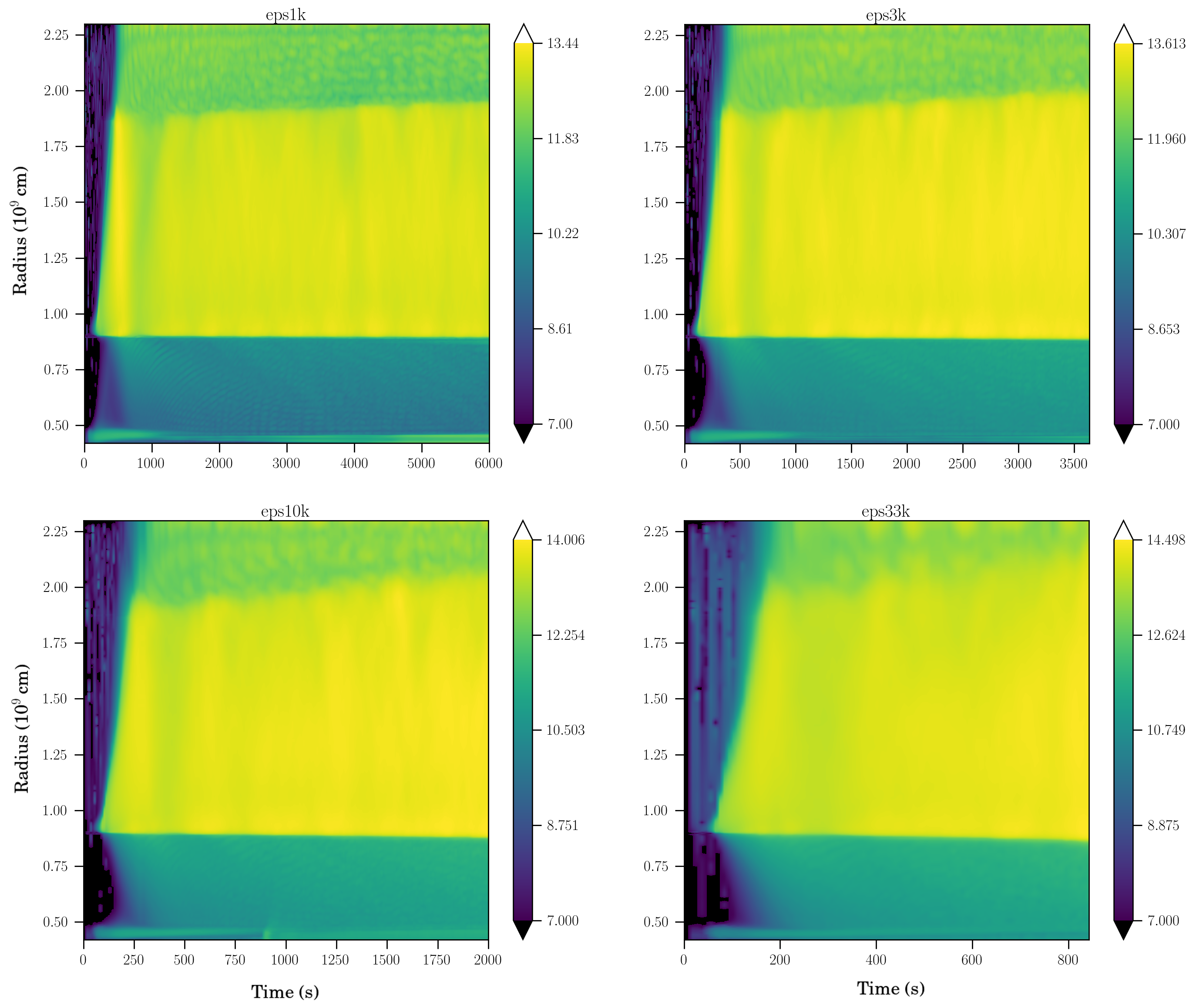} 
\caption{Contour plots of the logarithm of the total KE evolution in the radial direction for the \textsf{eps1k}, \textsf{eps3k}, \textsf{eps10k} and \textsf{eps33k} models. Each panel has the same radial scale (vertical axis) but has individual scales for the colour bar representing the KE magnitude and thus each model should be compared qualitatively and not quantitatively. Each model passes through an initial transient phase characterised by a strong pulse in KE near the start of the simulation. The models, clearly show extended periods of turbulent entrainment, characterised by the migration of the upper boundary into the stable region above, which can be seen from the radial extension of the relatively higher KE in the turbulent region (yellow region) compared with the upper stable region.}
\label{tke_contour}
\end{figure*} 

\indent In order to be computationally efficient while maximising the resolution, a mesh size of $512^3$ was chosen for all models. This resolution was shown to be sufficient to model the upper convective boundary for the \textsf{hrez} model of C17. \corrdre{Their \textsf{vhrez} model ($1024^3$) at $\epsilon_{fac}=10^3$ modelled the lower convective boundary more accurately, but \corr{it is currently} too computationally expensive to compute many models at such a resolution. With a resolution of 512$^3$, the effective Reynolds number of these simulations is Re$_{\rm eff}\sim$ 4000, placing the convective shells within the turbulent regime. See Appendix \ref{reeff} for a description of the effective Reynolds number.}\\
\indent Eight models are computed and are named according to the value used for $\epsilon_{fac}$, see Table \ref{eps_tab}. The model \textsf{eps1k} is an extension of the C17 \textsf{hrez} model up to 6000\,s. All other models were computed from the same initial conditions. This allows one to study the effect that varying $\epsilon_{fac}$ has on the initial transient stage. In particular, the time required to reach a quasi-steady state, where an equilibrium is reached between the driving luminosity and numerical dissipation is of interest. Figure \ref{eps_rho_entropy_log} shows the density (blue) and entropy (red) profiles for the 3D hydrodynamic model initial output and the corresponding 1D stellar model initial conditions (black). The highest energy model, \textsf{eps33k}, is difficult to evolve over long times as the energies are so extreme that the structure of the shell is disrupted, which results in the shell becoming dynamically unstable. As such, the physical simulation time for this model is relatively shorter than the other models. 
The global properties of each model are summarised in Table \ref{eps_tab}.\\

\begin{figure*}
\centering
\includegraphics[width=0.75\textwidth]{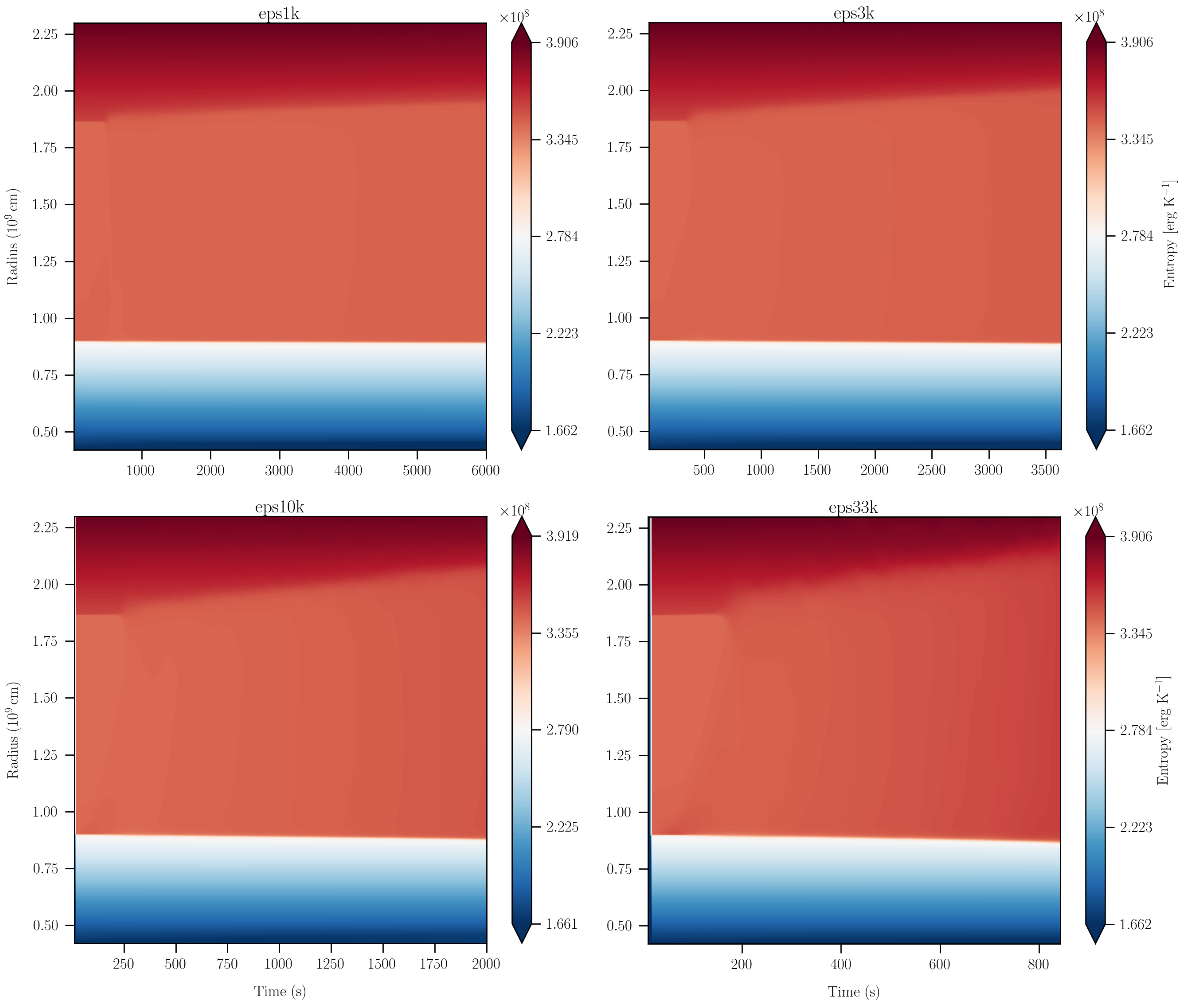}
\caption{\corrdre{Contour plots of the entropy in the radial direction for the \textsf{eps1k}, \textsf{eps3k}, \textsf{eps10k} and \textsf{eps33k} models. Each panel has the same radial scale (vertical axis) but has individual scales for the colour bar representing the entropy and thus each model should be compared qualitatively and not quantitatively.}}
\label{entropy_contour}
\end{figure*} 

\begin{figure*}
\centering
\includegraphics[width=1.05\textwidth]{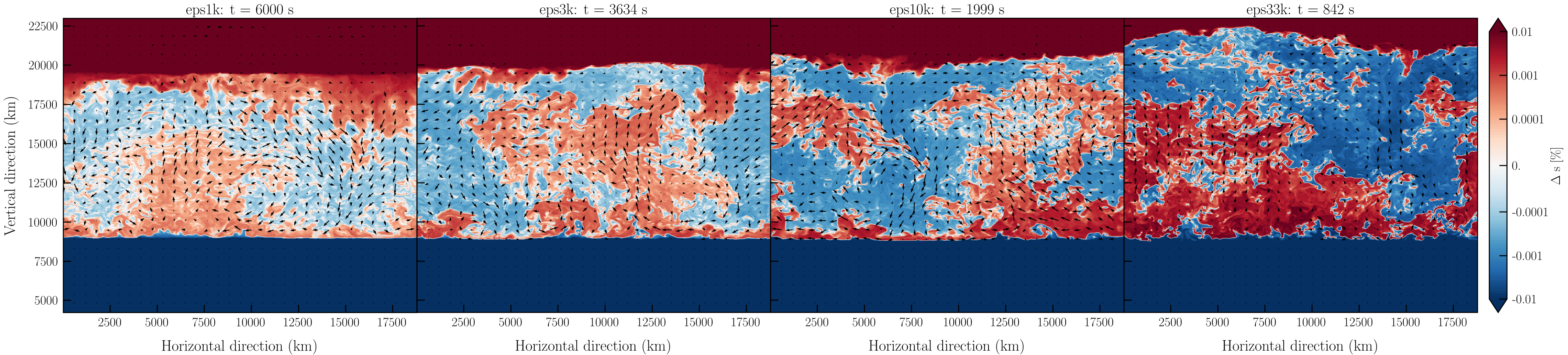}
\caption{\corrdre{Contour plots of the entropy variation from the mean entropy within the convective region. Models from left to right show the entropy variation of the \textsf{eps1k}, \textsf{eps3k}, \textsf{eps10k} and \textsf{eps33k} models at the final timestep of each model (the \textsf{eps33k} model is taken at 842\,s, before the shell is disrupted). The magnitude and direction of the arrows represent the speed and direction of the $v_x$ and $v_y$ velocity fields in the x-y plane.}}
\label{entropy_var}
\end{figure*} 

\section{Results and discussions}\label{eps_comp2}

\subsection{General flow properties}

\begin{figure}
\centering
\includegraphics[width=0.5\textwidth]{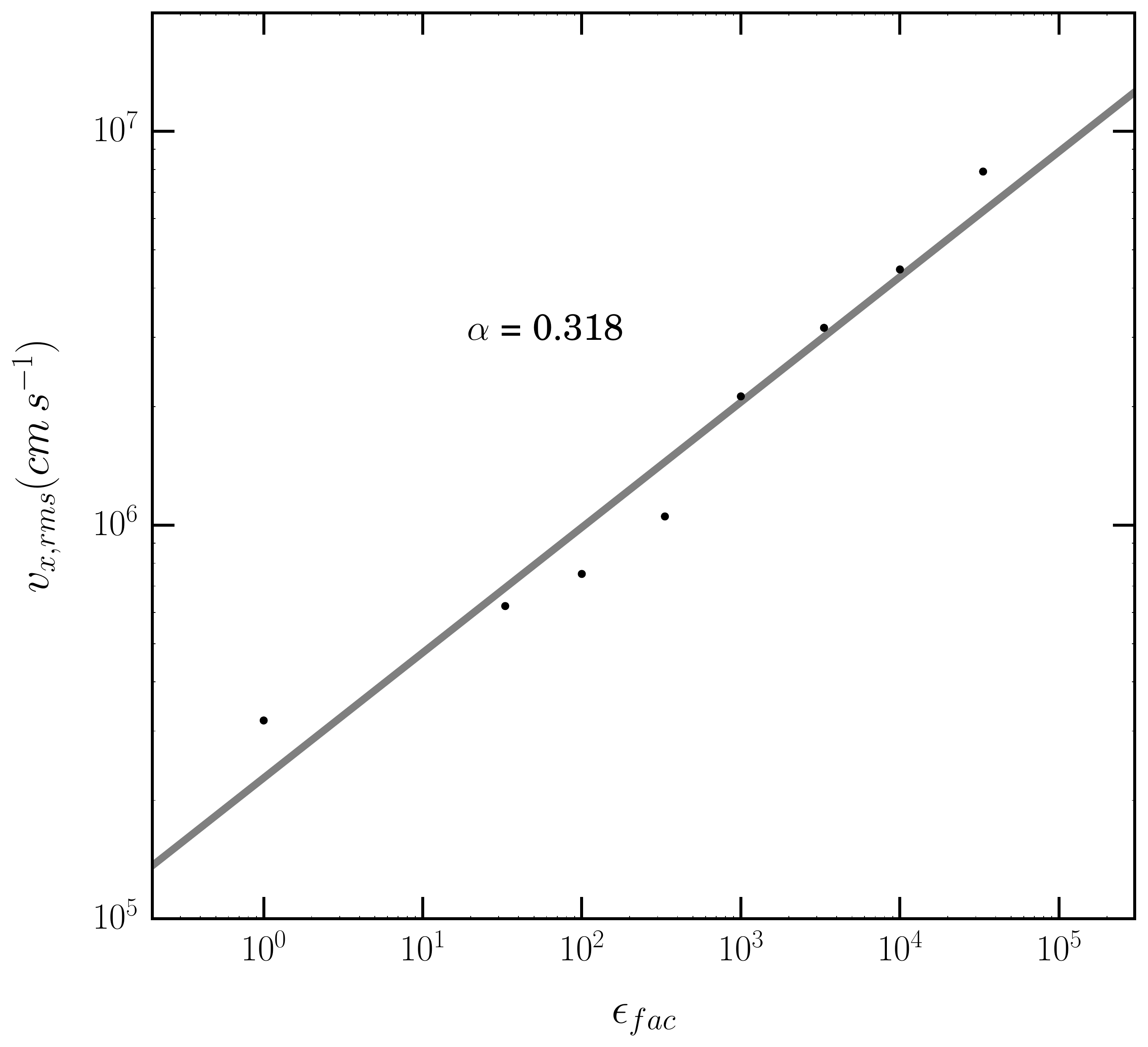} 
\caption[Radial velocity as a function of the boosting factor for the \textsf{eps1} - \textsf{eps33k} models]{Radial (x-direction) RMS velocity averaged over the convective zone versus boosting factor of the nuclear energy generation rate, for all models. This plot helps determine the scaling between these two quantities, assuming $v_{\rm x,rms}\propto \epsilon^{\;\alpha}$. A linear regression on this data provides a line of best fit with a slope of $\alpha=0.318\pm0.034$, implying that the vertical flow velocity is roughly proportional to the energy generation rate (or luminosity) to the power one third, or $v_{\rm x,rms}\propto L^{1/3}$.}
\label{vel_lum_scale}
\end{figure} 

\begin{figure*}
\centering
\includegraphics[width=0.75\textwidth]{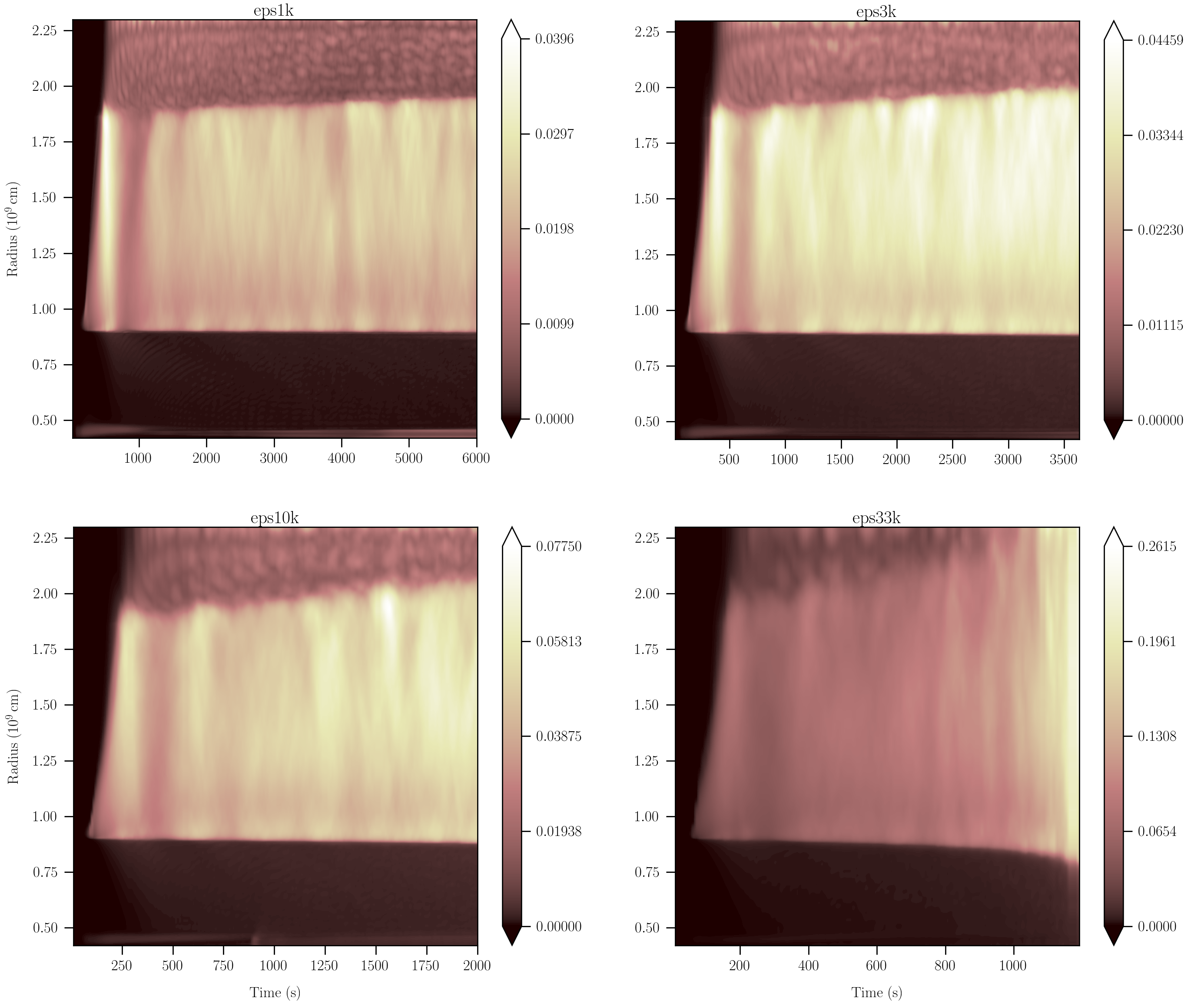}
\caption{\corrdre{Contour plots of the Mach number in the radial direction for the \textsf{eps1k}, \textsf{eps3k}, \textsf{eps10k} and \textsf{eps33k} models. Each panel has the same radial scale (vertical axis) but has individual scales for the colour bar representing the Mach number and thus each model should be compared qualitatively and not quantitatively. It should be noted that the entirety of the \textsf{eps33k} model has been included to show that toward the end of the model the fluid is approaching the transonic regime, where the Mach number $> 0.25$.}}
\label{mach_contour}
\end{figure*} 

\begin{figure}
\centering
\includegraphics[width=0.4\textwidth]{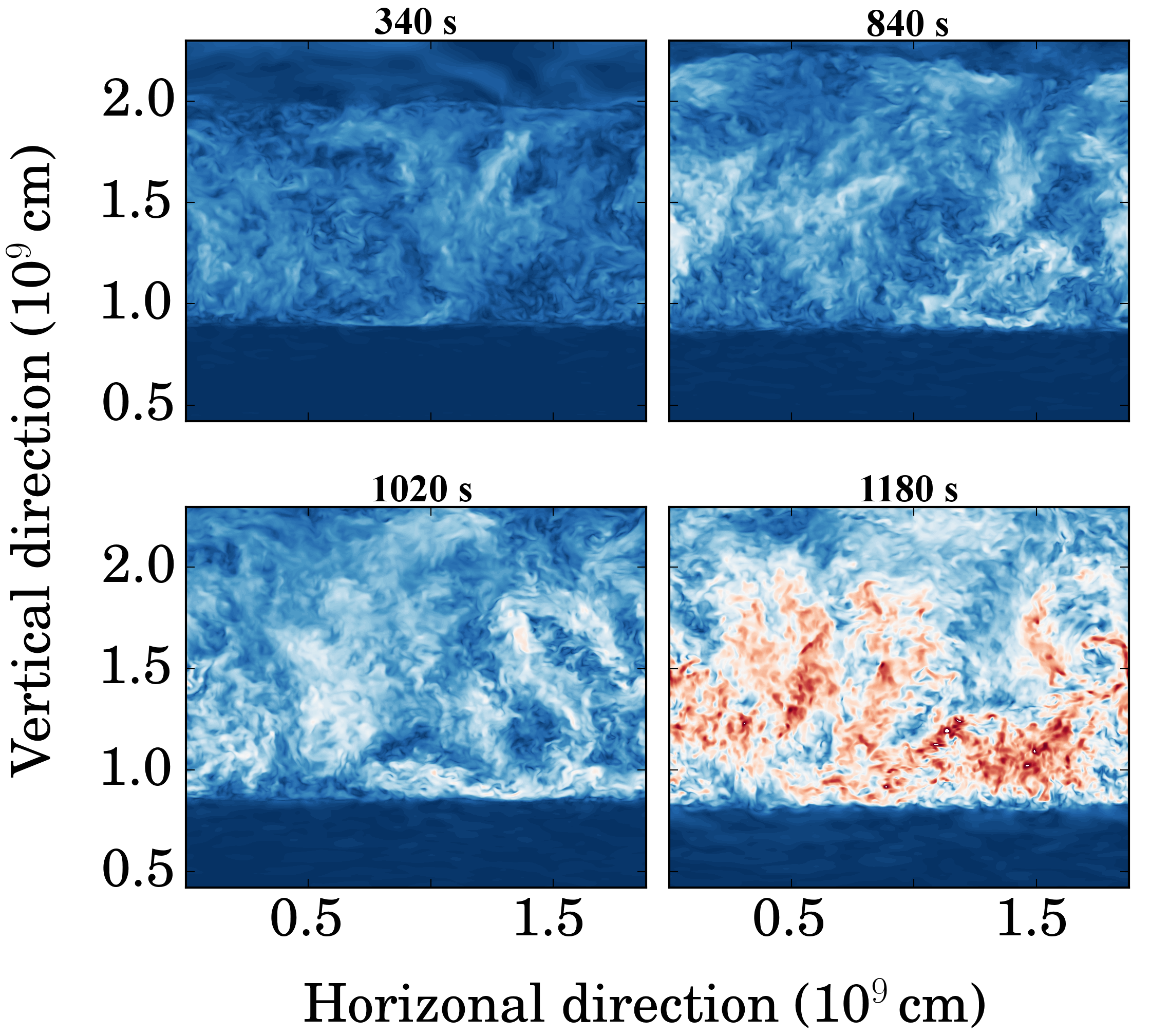}
\includegraphics[width=0.06\textwidth,height=0.375\textwidth]{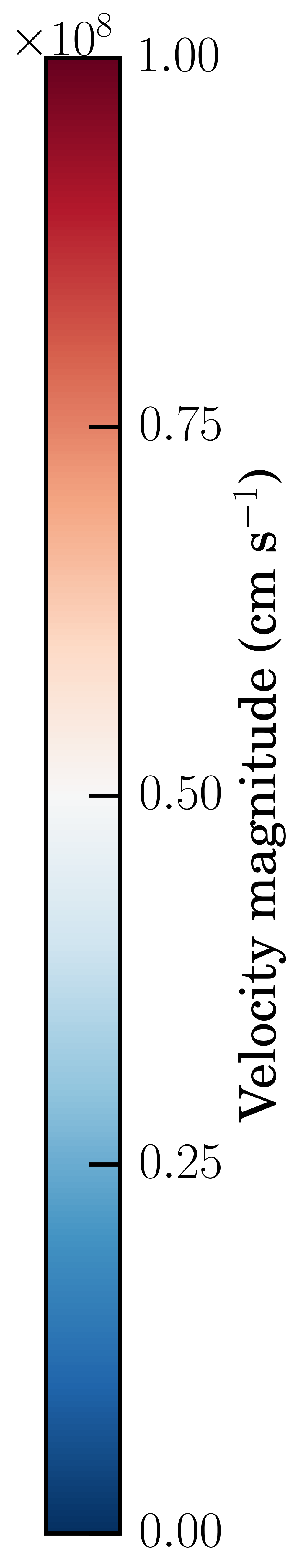}
\caption[Sequential velocity magnitudes in the $x-y$ plane of the \textsf{eps33k} model]{Sequential vertical cross-sections in the x-y plane of the velocity magnitude $\left(|v|=\sqrt{v_x^2+v_y^2+v_z^2}\,\right)$ for the \textsf{eps33k} model. Snapshots are taken at 340\,s (upper left), 840\,s (upper right), 1020\,s (lower left) and 1180\,s (lower right). The colour bar represents the values of the velocity magnitude in units of cm\,s$^{-1}$. The upper panels reveal the progressive expansion of the upper boundary layer into the surrounding stable region. The lower panels reveal that the upper boundary is then completely disrupted and the entire previously upper stable region becomes turbulent by the end of the simulations.}
\label{vmag_eps33k}
\end{figure} 

\indent The models first pass through an initial transient phase (whereby the turbulent velocity field is established) and the models eventually reach a quasi-steady turbulent state, where the initial conditions no longer influence the flow. The transition to the quasi-steady state for each model can be seen most easily in Fig. \ref{ke_and_ke-tke}, where the specific total kinetic energy (KE; thick solid lines), $\mathrm{KE_{tot}}=(v_x^2+v_y^2+v_z^2)/2$, for each model are plotted against the simulation time. The initial transient phase begins with a characteristic sharp rise in the KE up to a local maximum, which is then followed by a steady decrease. This marks the end of the initial transient, and the beginning of the quasi-steady state. From Fig. \ref{ke_and_ke-tke} it can be seen that the local maximum in KE during the transient phase increases with $\epsilon_{fac}$, and that the overall KE for each model increases as $\epsilon_{fac}$ is increased. This is intuitive as an increase in energy generation results in a larger flux of KE at the temperature peak near the bottom of the shell. \corrdre{The thin dotted lines in Fig. \ref{ke_and_ke-tke} show the specific total KE, $\mathrm{KE_{tot}}$ (thick solid lines) minus the specific total turbulent KE, $\mathrm{KE_{turb}}=(v_x'^{\,2}+v_y'^{\,2}+v_z'^{\,2})/2$ (see Appendix \ref{ssec:rans} for a description of the notation). These lines therefore represent the KE that is not associated with turbulent convection, i.e. KE due to waves propagating throughout the computational domain.} 

Specific KE spectra for each model are presented in Appendix \ref{app_spectra}.

\indent The \corr{logarithm of the} specific KE evolution over the radius of the computational domain for the \textsf{eps1k}, \textsf{eps3k}, \textsf{eps10k} and \textsf{eps33k} models are shown as contour plots in Fig. \ref{tke_contour}. The colour-bars show values of \corr{log(KE)} \corr{and are unique to each panel}. Each panel is labelled for the respective model. Each model passes through an initial transient phase characterised by a strong pulse in KE near the start of each simulation. The models show semi-regular pulses in KE over their evolution, a characteristic of convective transport. Such strong turbulent motions also lead to turbulent entrainment. This is best seen at the upper boundary by the gradual migration of this boundary into the stable region above. In all models, gravity waves (excited by the convective flows hitting the boundary) can be seen in the upper stable region, identified by short, horizontal \corr{yellow} streaks.

\indent \corrdre{The entropy evolution over the radius of the computational domain for the \textsf{eps1k}, \textsf{eps3k}, \textsf{eps10k} and \textsf{eps33k} models are also shown as contour plots in Fig. \ref{entropy_contour}. The colour-bars show values of the entropy in units of erg\,K$^{-1}$ \corr{and are unique to each panel}. Each panel is labelled for the respective model. }
Turbulent entrainment is visible via the increasing extent of the high-entropy convective region. 

\cdre{The panels in Fig. \ref{entropy_var} show contour plots for the final snapshot\footnote{\cdre{The \textsf{eps33k} model snapshot is taken at 842\,s rather than the final snapshot of the simulation, before the shell becomes disrupted.}} of the \textsf{eps1k}, \textsf{eps3k}, \textsf{eps10k} and \textsf{eps33k} models. Each contour plot shows the variation in entropy from the mean entropy in the convective region. This form of visualisation was chosen as the entropy in the convective region is nearly homogeneous. Entropy variations from the mean are seen in the convective region and at the boundaries, but are very small ($\sim10^{-4}$, c.f. fig. 6 of \citealt{2017MNRAS.471..279C}). The magnitude and direction of the arrows represent the speed and direction of the $v_x$ and $v_y$ velocity fields within the x-y plane. With an increase in boosting factor it can be seen that the variation in entropy is increasing within the convective region, shown by darker red and blue hues. The comparative global distortion of the convective boundaries, in particular the upper boundary, clearly increases with an increasing boosting factor, as would be expected. Smaller scale distortions along the boundaries, such as Kelvin-Helmholtz instabilities, are also more abundant with increasing boosting factor. }

Using dimensional analysis of a turbulent system, it can be shown that the energy dissipation rate is $\epsilon\sim v_{\rm rms}^3/\ell$, where $\ell$ is the integral length scale and $v_{\rm rms}$ represents the velocity of the largest energy-containing fluid elements \citep{1941DoSSR..30..301K}. As the energy dissipation rate is set by the energy generation rate which is proportional to the luminosity, it can be shown that\vspace{-2mm}

\begin{equation}\label{vel_lum}
v_{\rm x,rms}\propto L^{1/3},
\end{equation}
where $v_{\rm x,rms}$ is the vertical RMS velocity, assuming that the integral scale is constant and that $v_{\rm x,rms}\sim v_{\rm rms}$. Several other studies \citep{2000ApJS..127..159P,2009ApJ...690.1715A,2015ApJ...809...30A,2017MNRAS.465.2991J} have also confirmed this proportionality.\\
\indent For the eight models in this study, the radial (vertical) velocity does indeed have a positive correlation with the cube root of the nuclear energy generation rate. This is shown by Fig. \ref{vel_lum_scale}, where the vertical RMS velocity of each model is plotted against the boosting factor, $\epsilon_{fac}$. A linear regression on these values reveals a line of best fit with a slope of $0.318\pm0.034$. This is in agreement with Eq. \ref{vel_lum}.\\
\indent The time-scale of the transient phase and the time required to establish the turbulent velocity field is approximately one convective turnover time \citep[][C17]{2007ApJ...667..448M}. The above scaling implies that, as the energy generation rate is increased while the initial structure is kept constant, the convective turnover time (given in Table \ref{eps_tab}) will decrease. This is confirmed in Fig. \ref{ke_and_ke-tke} by the relatively shorter initial transient times and shorter fluctuation\footnote{Such fluctuations are associated with a phase lag between the buoyant driving and dissipative damping \citep{2011ApJ...741...33A}.} periods during the quasi-steady phase for the more energetic models.\\ 
\corrdre{From above, assuming that the integral length scale is constant, it can be shown that the Mach number scales with
 the nuclear energy generation rate. So that, for constant sound speed the Mach number varies with the boosting factor as Ma $\propto \epsilon_{fac}^{1/3}$. The temporal evolution of the Mach number over the radius of the computational domain is plotted in Fig. \ref{mach_contour} for models \textsf{eps1k}, \textsf{eps3k}, \textsf{eps10k} and \textsf{eps33k}. Again, each colour bar is unique to each model and shows the dimensionless Mach number. Similarly to Fig. \ref{tke_contour}, the episodic nature of turbulent convection and the generation of gravity mode waves in the upper stable region is well represented by the Mach number.}

\subsubsection{Model \textsf{eps33k}}\label{e33k}
Looking at Fig. \ref{tke_contour}, at first glance the \textsf{eps33k} model may appear to still be transitioning through the transient phase, but actually this transition is shown by the \corr{yellow} region spanning the convective region up to around 200\,s, followed by strong entrainment at the upper boundary. Towards the end of the model ($>700\,$s) a strong increase ($>10^{14}\,$erg\,g$^{-1}$) in KE can be seen. \corr{At that time}, the shell becomes dynamically unstable, and at later times ($>1000\,$s) the shell is completely disrupted. This can also be seen in the time series of the velocity magnitude snapshots for this model in Fig. \ref{vmag_eps33k}, where the velocity magnitude is plotted at 340\,s, 840\,s, 1020\,s and 1180\,s. Comparing the velocity magnitudes at 340\,s and 840\,s (top two panels), it can be seen that the upper boundary has migrated, almost entirely encompassing the stable region above. By 1020\,s (lower left panel) the upper boundary no longer exists and the entire previously upper stable region is now turbulent. These turbulent motions are no longer decelerated by approaching a stable region above, but still over-turn (as if these motions were approaching a boundary with an extremely high bulk Richardson number) due to the reflective boundary condition at the edge of the simulation domain. By 1180\,s, the removal of the upper stable region, leads to the turbulent velocities increasing dramatically (lower right panel) as turbulence is still driven by a large nuclear energy generation rate at the bottom of the shell, and the up-flowing radial velocities produced from this turbulence are not reduced by a negative buoyancy force due to the presence of an upper boundary. Instead the strong radial deceleration and turning of fluid elements at the upper domain boundary results in a band of material ($\geqslant 2\times10^9\,$cm) with visibly reduced velocities. Regardless, the removal of the boundary results in an un-damped, runaway acceleration within the convective region and the velocities become so high, that the fluid approaches the transonic regime ($|v|\sim10^8\,$cm\,s$^{-1}$) and the shell becomes dynamically unstable. \corrdre{This is shown in the bottom right panel of Fig. \ref{mach_contour}, where the Mach number can be seen to exceed $0.25$. Unsurprisingly, this demonstrates that for the given background stratification there is an upper limit to the amount of artificial boosting that can be applied to the model without complete shell disruption.} 
Despite such dynamic and violent behaviour, the short time period between $450\,$s - $900\,$s of the \textsf{eps33k} model still represents turbulent entrainment at a greatly boosted rate, this time window is used in further boundary analysis for this model.

\begin{figure}
\centering
\includegraphics[width=0.5\textwidth]{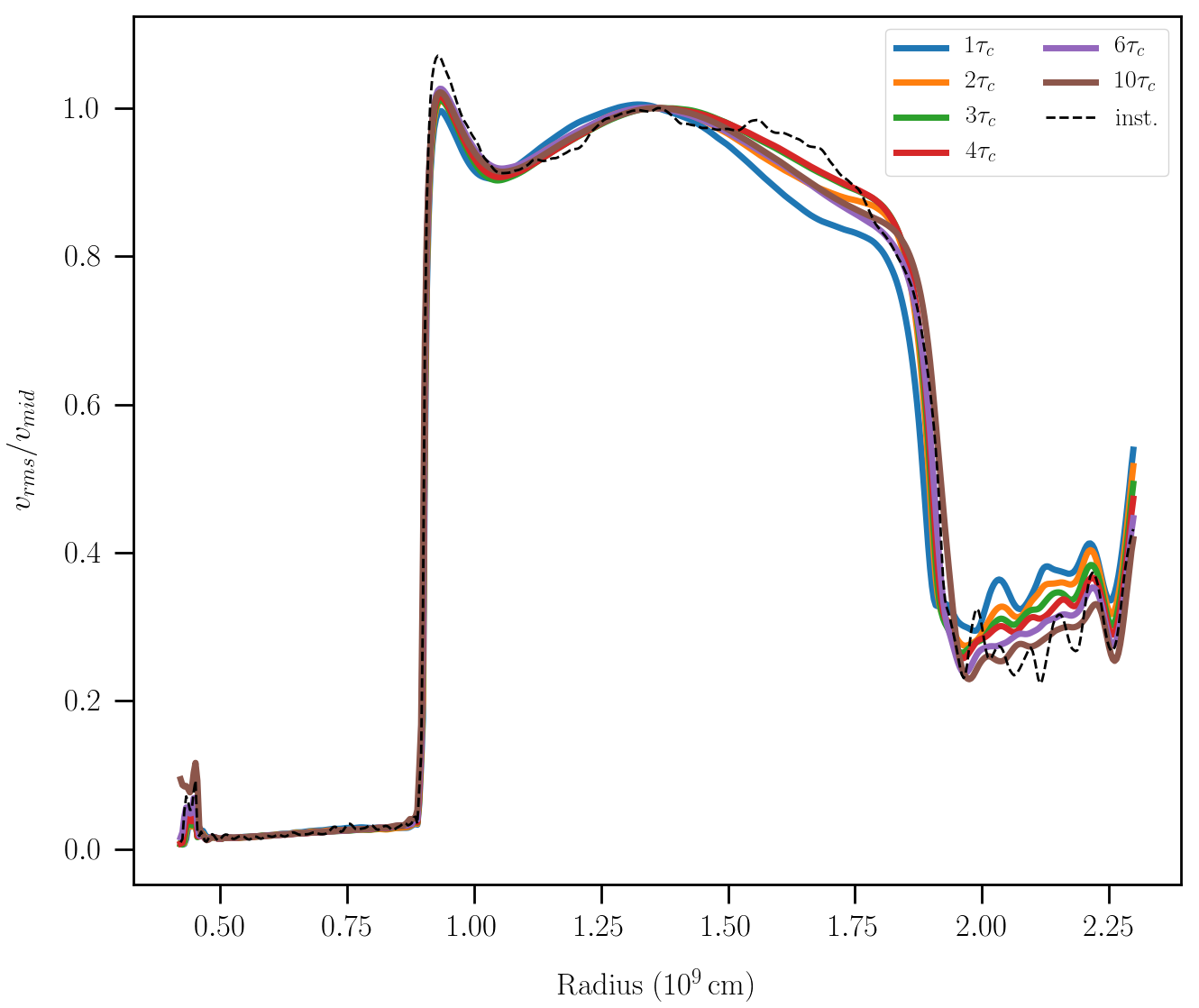}
\caption{\corrdre{Radial profiles of the RMS velocity normalised by the velocity in the centre of the computational domain. The coloured profiles show the velocity time averaged over different time windows of integer convective turnovers; these are one, two, three, four, six and ten convective turnovers for the \textsf{eps1k} model. The black dashed line shows the instantaneous velocity which has not been time averaged.}}
\label{vrms_time_check2}
\end{figure} 

\begin{figure}
\centering
\includegraphics[width=0.5\textwidth]{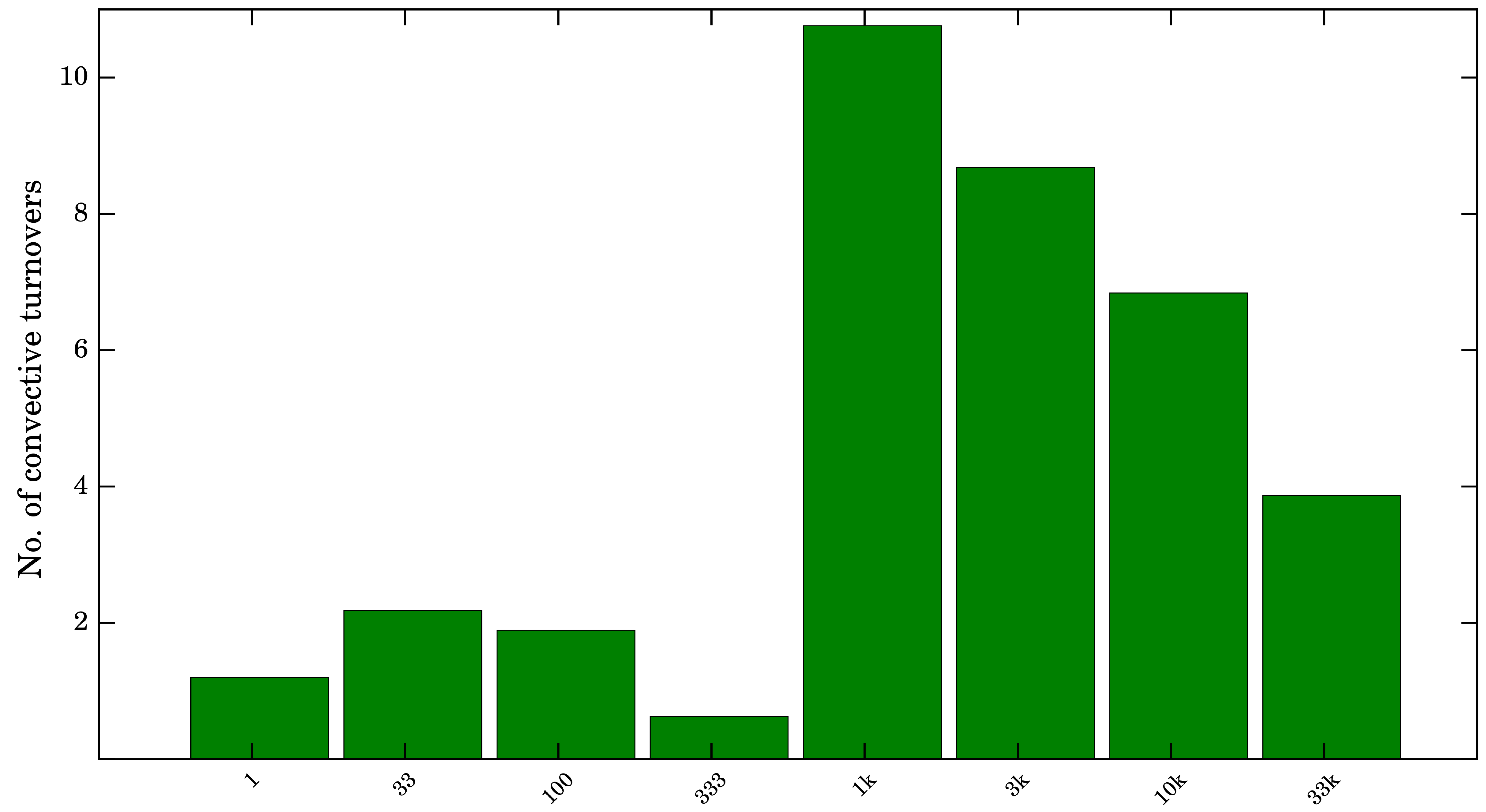}
\caption[Number of convective turnovers for the \textsf{eps1} - \textsf{eps33k} models]{Approximate number of convective turnovers following the initial transient for each model. Convective turnover times were calculated as the time taken for a convective eddy with speed $v_{\rm rms}$, to traverse twice the height of the convective region. The approximate time for entrainment to occur over a statistically significant period following the establishment of the turbulent velocity field is roughly between 3 and 5 convective turnover times. Hence, models \textsf{eps1, eps33, eps100} and \textsf{eps333} have not been evolved sufficiently, such that the turbulent region has been within a quasi-steady state for a significant time. The models which have evolved for one convective turnover or less are likely still adjusting to the initial conditions. Hence, only models \textsf{eps1k}, \textsf{eps3k}, \textsf{eps10k} and \textsf{eps33k} are included in the detailed boundary analysis.}
\label{turn_boost}
\end{figure} 
 
\subsection{Temporal evolution}
\indent \corrdre{The importance of including several convective turnovers for temporal averaging of quantities is demonstrated in Fig. \ref{vrms_time_check2}, where the RMS radial velocity profile (normalised to the velocity in the centre of the domain) is plotted using various temporal averaging windows from one to ten convective turnovers. An instantaneous profile (no time averaging) is also plotted as a black dashed line to illustrate the variance in velocity over a single snapshot in time. From Fig. \ref{vrms_time_check2} it can be seen that an averaging window of one convective turnover smoothes out the stochasticity of the instantaneous profile, but is qualitatively different from the remaining time windows.}

\indent The number of convective turnovers completed during the simulation time of each model is shown in Fig. \ref{turn_boost}, where it can be seen that the \textsf{eps1, eps33, eps100} and \textsf{eps333} models have all completed two or less convective turnovers. In such models, the turbulent velocity field may have developed but there is insufficient time during the quasi-steady state to provide statistically reliable results.\\
\indent Therefore, the four models \textsf{eps1}, \textsf{eps33}, \textsf{eps100} and \textsf{eps333} have not completed a sufficient amount of time in the quasi-steady phase to warrant a full analysis.
These models were thus excluded from the boundary analysis involving turbulent entrainment. \corrdre{It must be stressed that the poor temporal convergence of these models is not a result of the models themselves, but 
due to a lack of sufficient computational resources. We plan to evolve these models out to much later times ($> 10$ convective turnovers) in the future.}

\corrdre{The longer time-scales of the remaining models ($\gtrsim 4$ convective turnovers), leads to better converged properties over the longer quasi-steady states. An analysis of these models within the Reynolds-averaged Navier-Stokes framework and the effect that varying $\epsilon_{fac}$ has on the mean turbulent kinetic energy equation are presented in Appendix \ref{ssec:rans}.}\\
\indent \corrdre{The convective boundaries of these models are also in a quasi-equilibrated state, where the growth of the boundaries due to turbulent entrainment occurs within an equilibrium regime. This regime is known as the equilibrium entrainment regime \citep{2004JAtS...61..281F,2014AMS...1935.JGJG} and is achieved when the time-scale for boundary growth exceeds the convective turnover time-scale. In such conditions the boundary evolution is quasi-equilibrated as the entrainment process samples the entire spectrum of turbulent motions, rather than strong, individual plumes. The definitions for the two time-scales mentioned here and the respective values for the various models are discussed in Appendix \ref{equil_entrain}.}

\begin{figure*}
\centering
\includegraphics[width=0.75\textwidth]{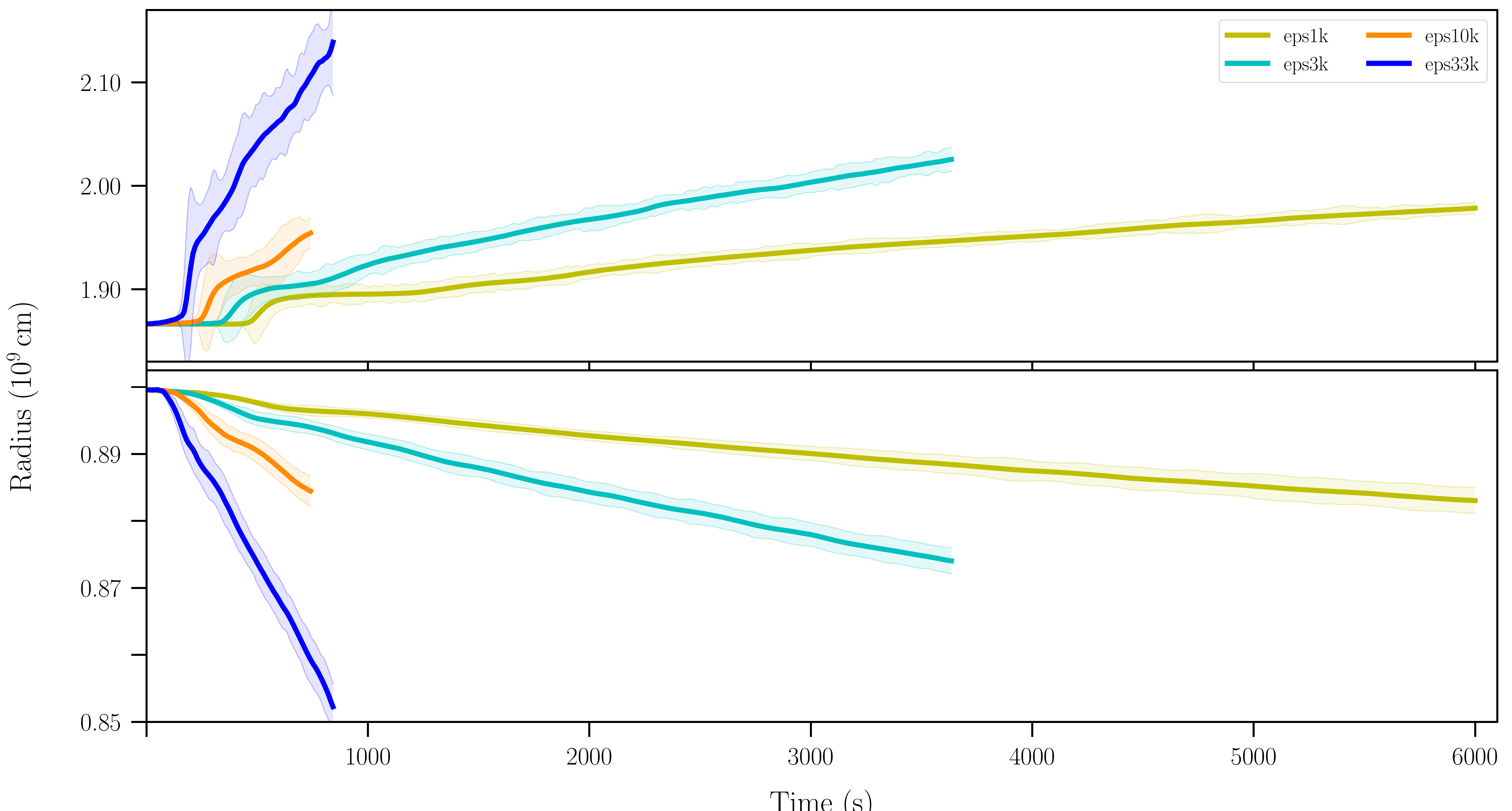}
\caption[Mean boundary position of the \textsf{eps1k} - \textsf{eps33k} models]{Time evolution of the mean radial position of the upper (top panel) and lower (bottom panel) convective boundaries, averaged over the 
horizontal plane for the \textsf{eps1k}, \textsf{eps3k}, \textsf{eps10k}, \textsf{eps33k} models. 
Shaded envelopes are twice the standard deviation from the mean boundary position. Variance in the mean boundary position generally increases with increasing driving luminosity, which can be associated with the occurrence of stronger and more frequent plume penetration of the boundary region. Note the different vertical scale between the upper and lower boundaries.}
\label{boundary_pos_reduced}
\end{figure*} 

\begin{figure*}
\centering
\includegraphics[width=0.49\textwidth]{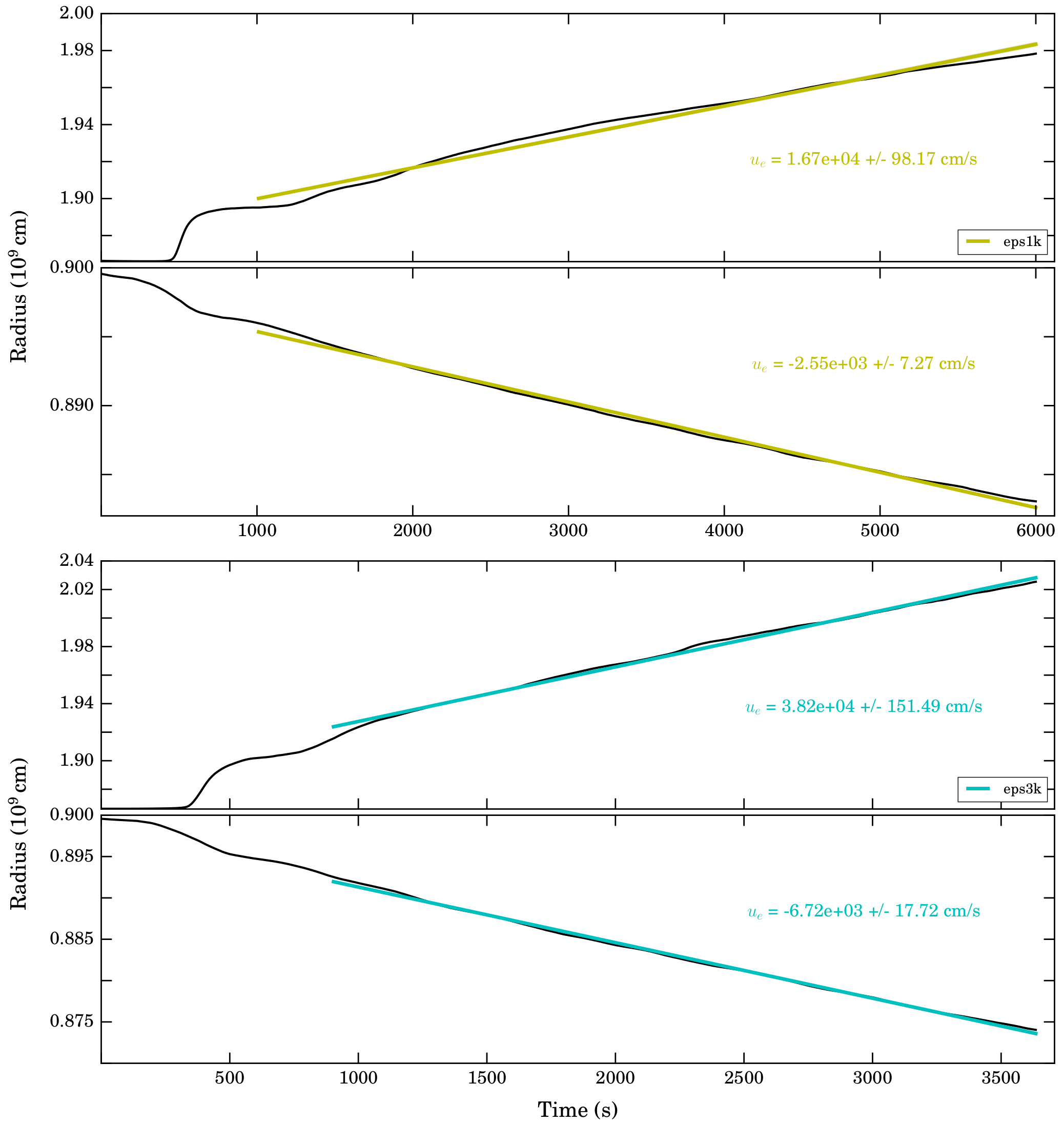}
\includegraphics[width=0.49\textwidth]{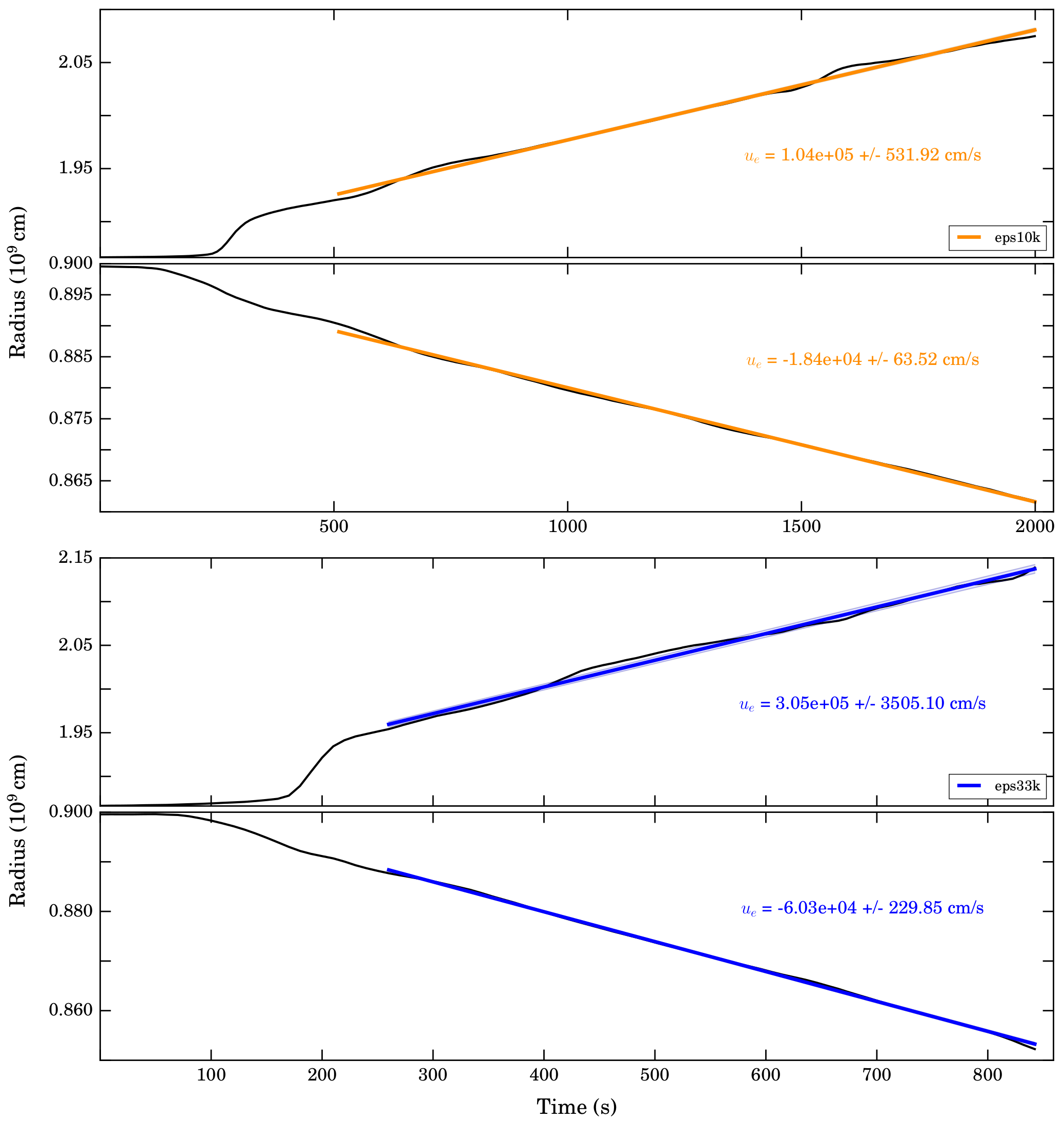}
\caption{Time evolution of the mean radial position of the upper (top panels) and lower (bottom panels) convective boundaries for the \textsf{eps1k} model (top left), \textsf{eps3k} model (bottom left), \textsf{eps10k} model (top right) and \textsf{eps33k} (bottom right). Note the different vertical scale between the upper and lower boundaries. Instantaneous boundary positions over the simulation times are shown by black lines, the coloured lines in each panel represent the best fit line following a linear regression of the boundary position over the quasi-steady state period. The corresponding entrainment velocity given by the best fit slope and respective error are given in each panel.}
\label{boundary_pos_ind}
\end{figure*} 

\subsection{Turbulent entrainment at convective boundaries}\label{entrain_eps}

In order to determine the positions of the convective boundaries, we adopt the method of C17, who calculate the positions based on the difference in composition between the adjacent convective and radiative regions. This provides a two-dimensional boundary surface. The estimated radial position of the boundary is then obtained through a horizontal mean of the radial position over this surface. These boundary positions over the simulated times of the \textsf{eps1k}, \textsf{eps3k}, \textsf{eps10k} and \textsf{eps33k} models are plotted in Fig. \ref{boundary_pos_reduced}. The positions of the upper (top panel) and lower (bottom panel) boundaries are shown by solid lines, and the shaded envelopes represent twice the standard deviation from the calculated means. The models show clear entrainment from the evolution of the boundary position over time.

The variance in the boundary positions increases as the driving luminosity is increased. This agrees with the notion that the boundaries become softer with an increased driving luminosity due to a higher TKE flux at the boundaries, and that a larger distortion in the boundary surface (relative to the mean) is characteristic of more plume penetration. 

The evolution of the convective boundary positions for the \textsf{eps1k} (top left), \textsf{eps3k} (bottom left), \textsf{eps10k} (top right) and \textsf{eps33k} (bottom right) models are also shown separately in Fig. \ref{boundary_pos_ind}. These plots are very similar to Fig. \ref{boundary_pos_reduced}, but allow an individual assessment of the entrainment and boundary migration for each model. The positions of the boundaries are shown by black solid lines. The best fit line from a linear regression of the boundary positions during the quasi-steady state for each model is shown by a solid coloured line. The corresponding best fit slope and relative error are shown in coloured text beside each fit. This slope is the entrainment velocity, and is remarkably close to linear in each case. The small errors in the best fit slope show this. The largest is 1.5\% for the upper boundary of the \textsf{eps33k} model.

\subsubsection{The entrainment law and boundary stiffness scaling}

\begin{figure}
\includegraphics[width=0.5\textwidth]{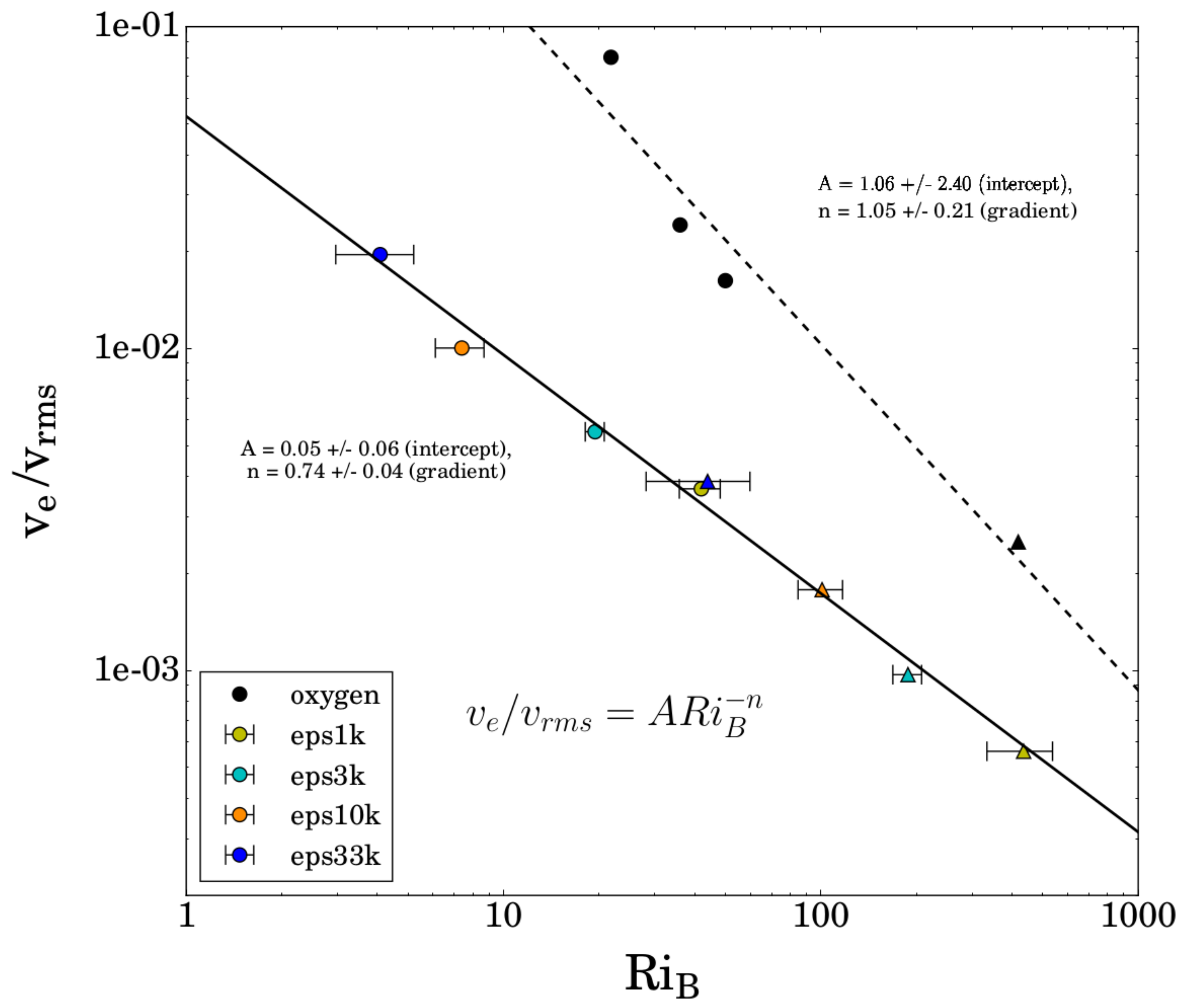}
\caption[Entrainment rate versus bulk Richardson number for the \textsf{eps1k} - \textsf{eps33k} models]{Logarithm of the entrainment speed (normalised by the RMS turbulent velocity) versus the bulk Richardson 
number. Coloured points represent data obtained in this luminosity study: \textsf{eps1k} (yellow); \textsf{eps3k} (cyan); \textsf{eps10k} (orange); and \textsf{eps33k} (blue). Triangles represent the values for the lower boundary and circles represent the values for the upper boundary. The horizontal error bars in the bulk Richardson number are the standard deviations from the mean over the quasi-steady state period. The black points represent data obtained in the oxygen burning study by \citet{2007ApJ...667..448M}. The solid and dashed lines show the best fit power laws to the 
respective data, obtained through linear regressions. The corresponding best fit slope and intercept along with the respective errors are shown for the current data (solid line) as $n=0.74\pm0.04$ and $A=0.05\pm0.06$, respectively, and for the oxygen shell (dashed line) as $n=1.05\pm0.21$ and $A=1.06\pm2.40$, respectively.}
\label{entr_law_eps}
\end{figure} 

The time rate of change of the boundary position through turbulent entrainment (the 
entrainment velocity), $v_e$, is known to have a power law dependence on the bulk Richardson number (Eq. \ref{rib}; see Appendix \ref{ssec:rib} for details). In this equation, the ratio of the entrainment velocity to the velocity representing the large-scale fluid elements, $E=v_e/v_{\rm rms}$, is considered \citep[e.\,g.][]{2014AMS...1935.JGJG}. This relationship between the relative entrainment rate and the bulk Richardson number is referred to throughout the meteorological and atmospheric science communities as the entrainment law, and is typically written as:

\begin{equation}\label{entr_law}
E=\frac{v_e}{v_{\rm rms}}=A\, \textrm{Ri}_{\rm B}^{-n},
\end{equation}

\vspace{2mm}where $A$ and $n$ are constants. Simulations (e.\,g. \citealt{1980BoLMe..18..495D}) and laboratory 
(e.\,g. \citealt{2010NPGeo..17..187C}) studies 
have found similar values for the coefficient, $A$, 
typically between 0.2 and 0.25. The exponent, $n$, is generally taken to be $1$. 
On the other hand, in a recent DNS study,
\citet{2013JFM...732..150J} showed that $A\approx 0.35$ and $n=1/2$.

\indent The \textsf{eps1k, eps3k, eps10k} and \textsf{eps33k} models have been interpreted within the framework of the entrainment law (Eq. \ref{entr_law}). These models cover a significant period ($\gtrsim4$ convective turnover times) in the quasi-steady state (see Fig. \ref{turn_boost}). The entrainment speed (normalised by the RMS turbulent velocity) is plotted as a function of the bulk Richardson number for these models in Fig. \ref{entr_law_eps}. The coloured circles and diamonds represent the values for the upper and lower boundaries, respectively, for the \textsf{eps1k} (yellow), \textsf{eps3k} (cyan), \textsf{eps10k} (orange) and \textsf{eps33k} (blue) models. 

The solid line is the best fit line following a linear regression of all the data points in (logarithmic) $E-\textrm{Ri}_B$ space. The line of best fit is then of the form $y=mx +c$, where $m=-n$ and $c=\textrm{log}A$. The slope ($n$) and intercept ($A$) along with the respective errors for this best fit are noted to the left of the line as $n=0.74\pm0.04$ and $A=0.05\pm0.06$, respectively. The black points in Fig. \ref{entr_law_eps} represent the entrainment rate, $E$ (as a function of the bulk Richardson number) obtained in the oxygen shell burning study by \citet{2007ApJ...667..448M}, and the dashed line is the best fit curve following a linear regression of their data points. The slope and intercept are also noted beside this fit as $n=1.05\pm0.21$ and $A=1.06\pm2.40$, respectively.

Comparing the values obtained for the constants in this luminosity study with those obtained in the resolution study by C17 (see their fig. 15), it can be seen that the values for $n$ are in reasonable agreement, and point towards a value in the range $1/2 \leq n \leq 1$ for fusion driven (neutrino cooling dominated) turbulent convection in the carbon  burning shell of massive stars. The value of $A$ is slightly smaller than the values quoted in other laboratory and numerical studies. The lower value of $A$ as compared with that of the oxygen shell burning simulation (black points) suggests that the efficiency of the work done by turbulent eddies on the stable stratification at the boundary of the carbon shell is less than that of the oxygen shell. \corr{This being said, the error in the calculation of $A$ for the oxygen shell is large, it stands to reason then that both fits could lead to the same value of $A$, within the error range of the simulations.}

It is not clear at this stage if a single set of parameters ($A$ and $n$) in the entrainment law would apply to all burning stages.
\corr{In order to reduce the errors in the calculation of both the bulk Richardson number and the entrainment rate,} models which span larger fractions of the quasi-steady state are \corr{required}. The variance of $n$ between the carbon and oxygen shells suggests that there may be additional mixing processes occurring besides entrainment in the carbon shell. Uncertainties in the `correct' value of $n$ for turbulent entrainment are also present throughout terrestrial simulations, where $n=1/2$ was found by \citet{2013JFM...732..150J} while $n=1$ was found by \citet{1991AnRFM..23..455F}, for example.

Furthermore, an approximate scaling relation can be obtained between the bulk Richardson number and the driving luminosity. This can then allow the determination of the stiffness of convective boundaries in 1D stellar models. This relation can be obtained by starting with the formula for the bulk Richardson number (Eq. \ref{rib}) and substituting for $v_{\rm rms}$ using Eq. \ref{vel_lum} leading to

\begin{equation}
\textrm{Ri}_B \propto v_{\rm rms}^{-2}\propto L^{-2/3},
\end{equation}
\label{rib_lum}

\begin{figure}
\includegraphics[width=0.5\textwidth]{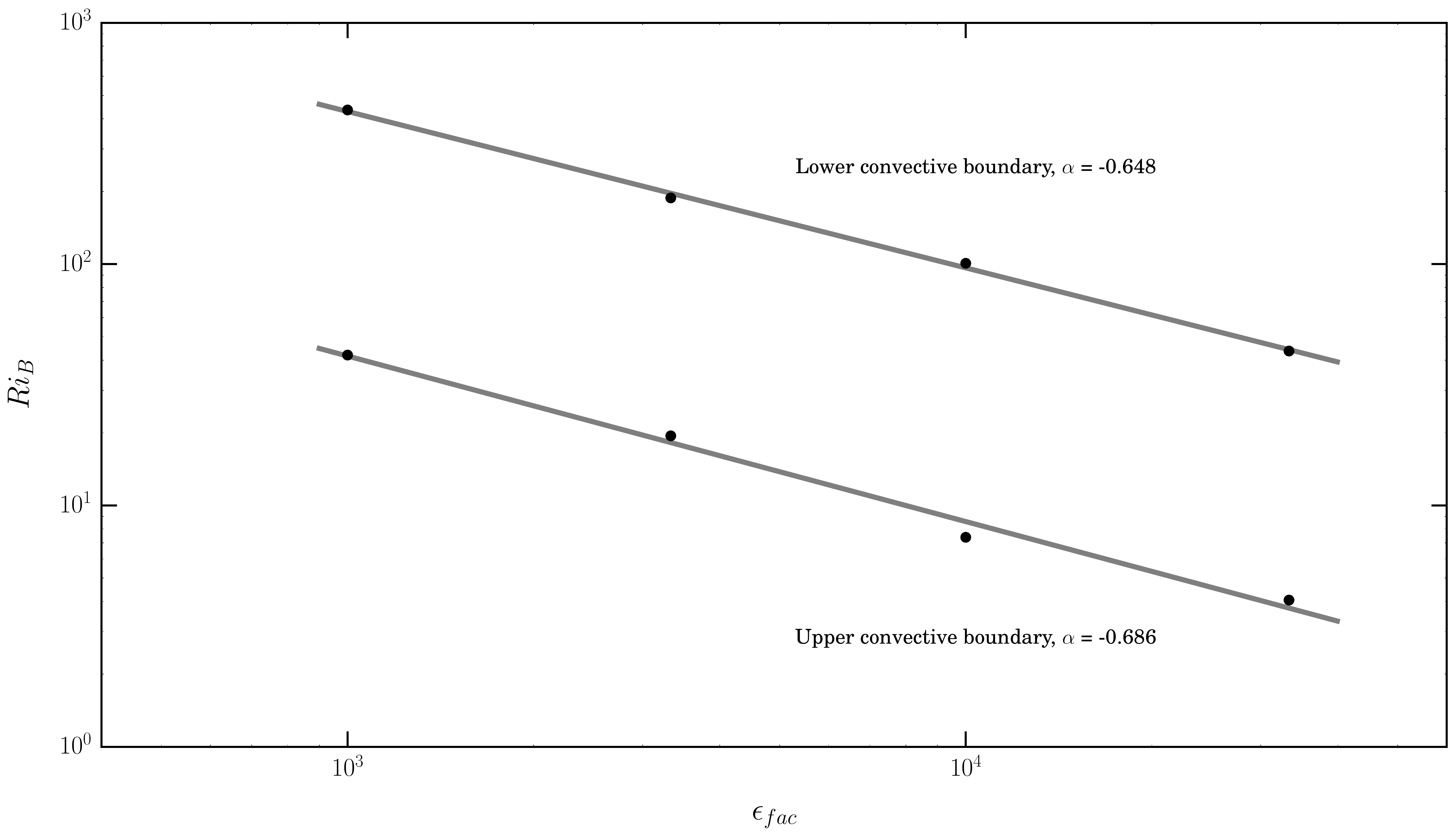}
\caption[Bulk Richardson number as a function of the boosting factor for the \textsf{eps1k} - \textsf{eps33k} models]{Bulk Richardson number as a function of the nuclear energy generation rate boosting factor. The upper points represent values for the lower convective boundary and the lower points represent values for the upper convective boundary for the \textsf{eps1k, eps3k, eps10k} and \textsf{eps33k} models. The grey lines are linear interpolations of each set of points, with the best fit slope noted next to each fit by $\beta$. In this form, the scaling between the boundary stiffness and driving luminosity is given by, Ri$_B\propto L^{\beta}$, where $\beta=-2/3 \;(-0.667)$ is the expected value.}
\label{rib_lum_plot}
\end{figure} 

\begin{figure*}
\includegraphics[width=0.75\textwidth]{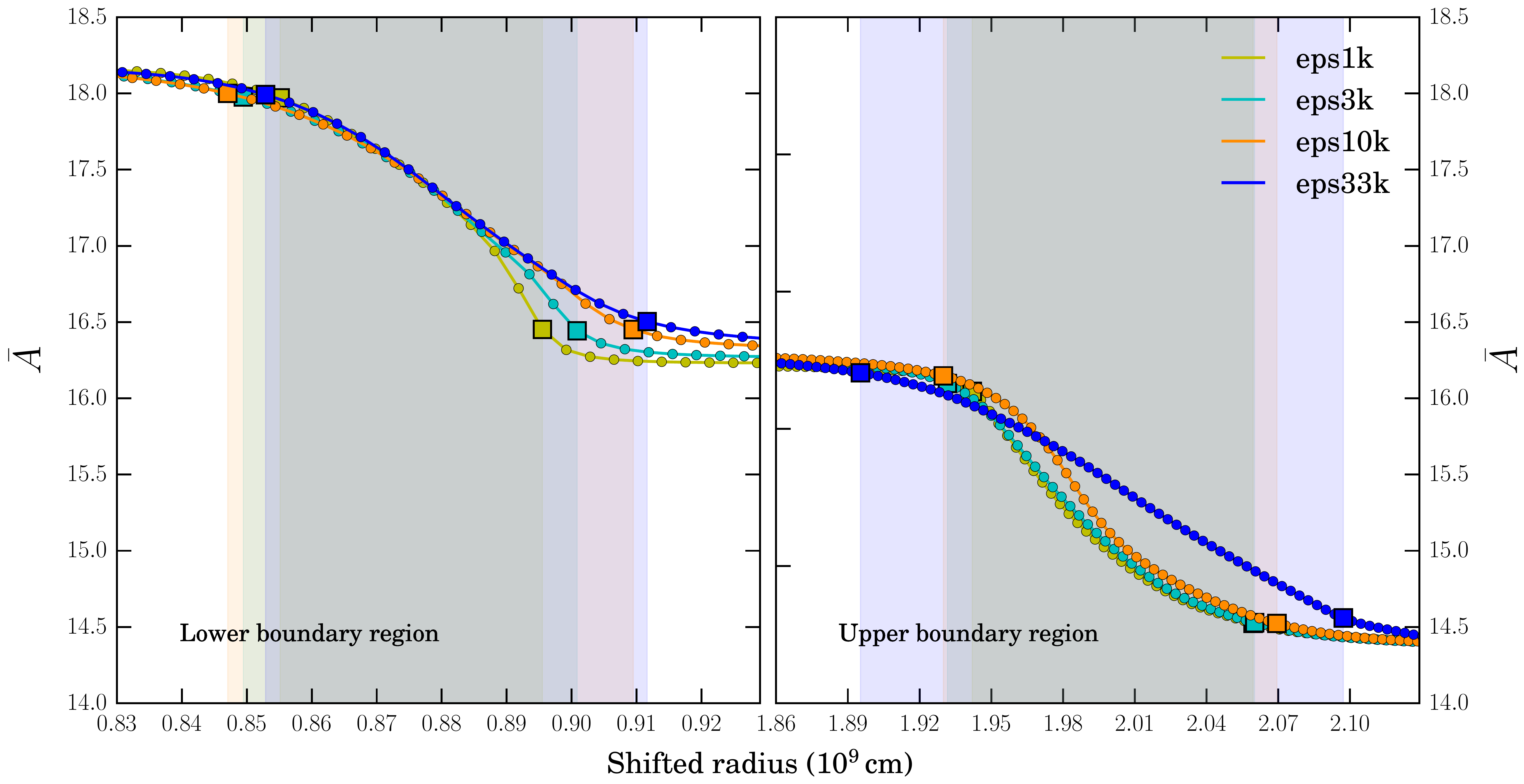}
\caption{Radial compositional (mean atomic mass) profiles at the lower (\textsf{left}) and upper (\textsf{right}) convective boundary regions for the last time step of the: \textsf{eps1k} (yellow); \textsf{eps3k} (cyan); \textsf{eps10k} (orange); and \textsf{eps33k} (blue) models. 
The radius of each model is shifted such that the boundary position coincides with that of the \textsf{eps1k} model. 
This makes it easier to compare the slope of the boundary at the final time-step for the different driving luminosities. Individual 
mesh points are denoted by filled circles. Approximate boundary extent (width) is indicated by the distance between two filled squares for 
each model, and is shown by the correspondingly coloured shaded regions. See \S4.5.3 of C17 for details on the definition of the boundary width.}
\label{eps_width}
\end{figure*} 

\noindent if the integral scale and buoyancy jump ($\Delta B$) are assumed to be constant (reasonable assumptions given an initial hydrostatic stratification and short dynamical time-scales).

This scaling between the driving luminosity and bulk Richardson number can easily be tested within the current luminosity study. As previously mentioned, only the four models \textsf{eps1k} - \textsf{eps33k} are used to test this relation, as the remaining models do not contain a sufficient number of convective turnovers over which entrainment can occur. The bulk Richardson numbers for the upper (bottom line) and lower (top line) boundaries of these four models are plotted in Fig. \ref{rib_lum_plot} as a function of the boosting factor, $\epsilon_{fac}$. A linear regression of the data points for each boundary is performed over logarithmic space in order to determine a best fit power law and scaling exponent, $\beta$, assuming Ri$_B \propto L^\beta$. This best fit power law is shown by the grey line for each boundary and the corresponding value of the scaling exponent is noted beside each one, $\beta=-0.686$ for the upper boundary and $\beta=-0.648$ for the lower boundary. For both boundaries, the value obtained is close to the expected value of $\beta=-0.667$. Unfortunately, due to such a sparse data set it is not possible to obtain errors in the values of the slope. Nevertheless, such an agreement between theory and simulation is encouraging and with more data points and longer time sampling for each model the confidence in these estimates can be improved. 

From the best fit power laws calculated in Fig. \ref{rib_lum_plot}, the bulk Richardson numbers can be extrapolated back to the original carbon burning luminosities of the 1D input model. These are Ri$_{B,u}=4750$ and Ri$_{B,l}=3.77\times10^4$ for the upper and lower boundaries, respectively. These values are in agreement (to within a factor of around 3) with the values calculated in the \textsf{eps1} model (see Table \ref{eps_tab}), although arguably, the \textsf{eps1} model values are not well constrained either due to the short number of convective turnovers simulated (e.g. see Fig. \ref{turn_boost}). The above values also agree (also to within a factor of around 3) with the values calculated from the stellar model initial conditions (see table A1 of C17), which also have inherent uncertainties, due to the difficulty of calculating the bulk Richardson numbers over coarse stellar model grids. 

The entrainment velocities for the \textsf{eps1} model can also be estimated using the above values for the bulk Richardson numbers. Inserting the above values for the bulk Richardson numbers into the entrainment law (Eq. \ref{entr_law}) with the values for $A$ and $n$ determined in Fig. \ref{entr_law_eps} and the RMS velocity from Table \ref{eps_tab} yields entrainment velocities of around 14\,cm\,s$^{-1}$ and 65\,cm\,s$^{-1}$ for the lower and upper boundaries, respectively. \corr{The corresponding mass entrainment rates can be estimated using, $\dot{m}_e=4\pi r^2\rho\,v_e$, where the density is that of the boundary region averaged over the quasi-steady state. The mass entrainment rates for the lower and upper boundaries of the \textsf{eps1} model are $1.35\times10^{-8}$\,M$_\odot\,$s$^{-1}$ and $3.77\times10^{-8}$\,M$_\odot\,$s$^{-1}$, respectively. Such mass entrainment rates imply that  the time required for a mass comparable to that of the carbon shell ($\sim0.75\,$M$_\odot$) to be entrained is roughly half a year. 
This time-scale seems rather short though considering that the evolutionary time-scale for the carbon shell used as input is roughly 30 years, suggesting that the inferred mass entrainment rate is too large compared to what is expected from stellar evolution models.}\\ 
\indent {It is very important at this point to remember that the initial 1D model used in this study corresponds to the start of the carbon shell burning phase, during which the convective region also grows significantly in the 1D calculation. We do not expect the mass entrainment rate derived in this study to apply for the entire carbon burning stage nor to other burning stages. The Bulk Richardson number is expected to evolve during a burning stage and between burning stages. It is expected to increase significantly as the entropy and mean molecular weight gradients at the convective boundaries build up throughout the burning stage. At the very end of a burning stage, the driving luminosity will decrease and entrainment is expected to cease altogether. { This is why a theoretical framework such as the entrainment law is crucial in order to correctly apply results of 3D hydrodynamic simulations back into 1D stellar evolution models} and to avoid unrealistic growths of convective zones.}\\
\indent Despite the difficulties discussed above, the reliable estimation of the bulk Richardson number is still best pursued through long time-scale (ideally $\gtrsim10$ convective turnovers), 3D hydrodynamic simulations of turbulent convection. The results of which can populate the entrainment law parameter space ($E-\textrm{Ri}_B$) and constrain the free parameters $A$ and $n$. The agreement with the scaling relation, Ri$_B\propto L^{-2/3}$, for several models over a range of different driving luminosities, demonstrates the applicability of this scaling relation to stellar flows. The discrepancy between the extrapolated values of the bulk Richardson number for natural carbon burning luminosities and the stellar model initial conditions, could be used to fit free parameters within the formulation of the bulk Richardson number for stellar models, namely the integration length, $\Delta r$, and the integral scale, $\ell$ (see Appendix \ref{ssec:rib}).\\
\indent With better constrained values of $A$ and $n$ as well as more accurate estimates of the bulk Richardson numbers within stellar models, new mixing prescriptions can be incorporated into stellar models, such as the one suggested by \citet{2007ApJ...667..448M}
\begin{equation}
\dot{m}_e=\frac{\partial m}{\partial r}\,v_e=4\pi r^2\rho\, v_{\rm rms}\,A\,\textrm{Ri}_B^{\,-n}.
\end{equation}

\subsection{Effects of driving luminosity on boundary widths}

\indent The composition profiles at both boundaries for the final time-step of each of the four models (\textsf{eps1k - eps33k}) are shown in Fig. \ref{eps_width}. Each composition profile has been shifted in radius such that the boundary position coincides with that of the \textsf{eps1k} model (for ease of comparison). Boundary region widths are calculated through the adoption of the method used by C17 (see their \S4.5.3). \corr{This method uses the jump in composition, $\bar{A}$, between the convective and stable regions. The boundary region is estimated to include all but 1\% of the composition contained within the convective region and surrounding stable region.}\\
\indent In Fig. \ref{eps_width}, the boundary width is denoted for each model by the distance between two filled squares at the edges of the boundary region, and is shown by the respectively coloured shaded region. These boundary widths and their fractions of the pressure scale height are given in Table \ref{tabbw_eps}.\\
\indent For the upper boundary (right panel of Fig. \ref{eps_width}) the composition profiles are similar except for the \textsf{eps33k} model, which has a shallower slope over the boundary region. In this model the upper boundary is much broader, and the increased driving leads to a dramatic smoothing in the abundance slope; this is not seen in the other models. As noted in \S\ref{e33k}, the strong increase in driving luminosity, is likely leading to a structural readjustment of the shell at this time-step and will eventually result in the complete disruption of the shell.\\\\
\indent 
At the lower boundary (left panel) the abundance profiles are very similar in slope between all of the different cases, with an increase in driving luminosity leading to a clear increase in boundary width. As entrainment becomes more effective with stronger driving luminosity, the composition in the convective region near the boundary also increases slightly. The increased amount of heavier material is due to mixing of material from the stable region below into the turbulent region.\\
\indent The smoothing of the horizontally averaged abundance profile due to the boundary deformation can be measured by the standard deviation of the boundary position (Fig. \ref{boundary_pos_reduced}). The deviation can be associated with the vertical fluctuations in the boundary surface, possibly due to plume penetration events. The extent of the boundary region (its width) can be associated with the strength of shear mixing (or Kelvin-Helmholtz instability) at the boundary caused by the U-turning of turbulent fluid elements.
The boundary widths (Table \ref{tabbw_eps}) have a weaker dependence on the boosting factor, $\epsilon_{fac}$, than originally expected, \corrdre{but it can be said that the boundary width generally increases with boosting factor, leading to 
a better spatial sampling of
the boundaries. The eps33k model with a boosting factor of $3.3\times10^4$ is the only model where both convective boundaries are spatially resolved at a resolution of $512^3$ (see the bottom left panel of Fig. \ref{mfa_ke_1k_3k} and Fig. \ref{eps_eku}). This is encouraging as it implies that there exists a resolution where the ILES method adopted here can provide a spatially resolved model of the carbon shell and its boundaries at nominal luminosity.}

\begin{table}
\centering
\begin{tabulary}{0.5\textwidth}{l || c c c c}
\hline\hline\\
 & \textsf{eps1k} & \textsf{eps3k} & \textsf{eps10k} & \textsf{eps33k}\\\\
\hline \hline\\
\textsf{$2\sigma_{r,l} (\rm cm)$}  &  3.88$\times10^6$  &  3.88$\times10^6$ & 5.96$\times10^6$ & 6.77$\times10^6$\\\\
\textsf{$2\sigma_{r,u} (\rm cm)$}  &  1.16$\times10^7$  &  2.26$\times10^7$ & 2.25$\times10^7$ & 1.04$\times10^8$\\\\
\textsf{$\delta r_l (\rm cm)$}  &  4.04$\times10^7$ & 5.14$\times10^7$ & 6.24$\times10^7$ & 5.88$\times10^7$\\\\
\textsf{$\delta r_u (\rm cm)$}  &  1.18$\times10^8$  &  1.29$\times10^8$ & 1.40$\times10^8$ & 2.02$\times10^8$\\\\
\textsf{$\delta r_l/H_{p,l}$} & 0.141 & 0.181 & 0.218 & 0.201\\\\
\textsf{$\delta r_u/H_{p,u}$} & 0.339 & 0.366 & 0.389 & 0.535\\\\
\hline\hline
\end{tabulary}
\caption[Boundary widths for the \textsf{eps1k} - \textsf{eps33k} models]{\corr{$2\sigma_{r,l}$ and $2\sigma_{r,u}$ are twice the standard deviations from the horizontal means of the boundary surface positions (shaded envelopes in Fig. \ref{boundary_pos_reduced}). These measurements can be associated with boundary surface height fluctuations, possibly due to plume penetration events.} $\delta r_l$ and $\delta r_u$ are the approximate boundary widths determined from the 
composition profiles of the lower and upper convective
boundaries respectively (see Fig. \ref{eps_width}). These measurements of the boundary represent the size of the boundary region which is formed due to  shear mixing at the boundaries. $H_{p,l}$ and $H_{p,u}$ are the average pressure scale heights across 
the lower and upper convective boundary regions, respectively.}
\label{tabbw_eps}
\end{table}

\vspace{3cm}
\section{Conclusions}\label{conc}
In this paper, we studied the dependence of convective boundaries on their stiffness and the turbulence strength. To study this dependence in a controlled way, we performed a series of simulations of the carbon burning shell, in which the nuclear energy generation rate was artificially boosted. The initial conditions were taken from the 15\,M$_\odot$ stellar model described in C17. Each model was computed within a Cartesian cube consisting of 512$^3$ computational zones. The nuclear energy generation rate was calculated using the same prescription described in C17, with the addition of a boosting factor, $\epsilon_{fac}$, which we varied between 1 and $3\times10^4$. The results showed that over the short dynamical time-scales, which these models were run over (due to limited computational resources), the shell structure was not adversely affected by the boosting (except in the most extreme case) and served to accelerate the evolution of the shell through increased rates of entrainment and mixing. As the structure was unchanged between each model, the vigour of turbulence was also increased for the same stable stratification and so the representation of entrainment in the most energetic cases ($\epsilon_{fac}\geq 10^4$) was far from been physically realistic during the adopted evolutionary state.\\ 
\indent Such a luminosity study allowed us to populate the parameter space ($E-\textrm{Ri}_B$) of the entrainment law (Eq. \ref{entr_law}) and constrain the values of $A$ and $n$ for high P\'{e}clet number, fusion driven convection. We found values of $A=0.05\pm0.06$ and $n=0.74\pm0.04$. The central result of this study is the strikingly clear dependence of the entrainment rate on the stiffness of the convective boundary. This result was found for both the top and bottom boundaries. The bottom boundary being stiffer than the top one, entrainment at the bottom boundary is slower. While this was already observed in our previous study C17, the present luminosity study places this finding on a firm footing.

Comparing to the values found in the resolution study by C17, a similar value for $n$ was found, and further points to a value of $n$ for fusion driven (neutrino cooling dominated) turbulent convection in the range $1/2 \leq n \leq 1$. The variance of $n$ between the carbon and oxygen burning shells \citep{2007ApJ...667..448M} suggests that there may be additional mixing processes occurring besides entrainment in the carbon shell. 
Comparison to the simulations of \cite{2013JFM...732..150J} suggests that these entrainment values may depend upon the P\'eclet number. Further, the structure of the convective region changes as evolution proceeds.  
While the entrainment law seems valid for times of the order of the turnover time, it may not be universal for evolutionary time scales and conditions.

Further simulations of stellar convection are needed to help populate the entrainment law parameter space. Longer time-scale simulations ($>10\,\tau_c$) are also needed to provide better statistics during the quasi-steady state and reduce the errors in the calculated values of $A$ and $n$. As computing power increases, and the porting of existing codes to GPU technologies is becoming more common, more high-resolution, longer time-scale simulations will be possible.

\indent This luminosity study also confirmed the scaling relation between the vertical RMS velocity and the driving luminosity, $v_{\rm x,rms}\propto L^{1/3}$, as found in previous studies. This study also confirmed the expected scaling between the bulk Richardson number (stiffness) of the boundary and the driving luminosity of the shell, Ri$_B\propto L^{-2/3}$. Such a relation will prove useful in developing new CBM prescriptions for entrainment in stellar evolution models. For example, it helps to extrapolate values of the bulk Richardson number back to the nominal carbon burning luminosities. Comparing these extrapolated values to those from the initial 1D stellar evolution model calculations in C17 reveals that they are similar to within a factor of around 3.\\
\indent This study represents a few steps along the very long road of integrating the results of 3D hydrodynamics simulations into a theoretical framework that can be used in 1D stellar evolution models; framework coined ``321D'' by John Lattanzio and others. Our simulations give a plausible representation for the physics of turbulent stellar convection, but only for a few turnover times.  Longer term behaviour, which is crucial for stellar evolution properties such as the size of mixing regions, remains a challenge. The evolution of the convective region may modify the entrainment law, which would depend upon the properties of the convection zone (e.g., luminosity, entropy jumps, composition jumps, etc.).
The problem may be a coupled one, so that there might not be a universal entrainment law as suggested by the different values derived for $A$ and $n$ between carbon and oxygen burning. Despite this potential non-universality, 3D simulations at various points throughout the evolution of stars may still be able to constrain the values of $A$ and $n$ to be used with the entrainment law in 1D models.\\
\indent
Including entrainment (or the penetrative overshoot commonly used in GENEC and other 1D models) moves the convective boundaries but does not change their shape. As shown in Fig.\,\ref{eps_width}, abundance profiles at convective boundaries in 3D simulations are smooth, sigmoid-like functions rather than the step-like functions found in GENEC and other 1D codes. These smoother shapes are due to convection-induced shear mixing (probably Kelvin-Helmholtz instability) occurring at the boundary as the fast upward convective flow U-turns. Note that rotation-induced shear mixing \citep{2017A&A...604A..25E} plays a similar role as the convection-induced shear and both lead to smoother profiles at boundaries.
Using a diffusion coefficient, which is exponentially decaying from the convective boundary in 1D codes following multi-D simulations of outer convective regions \citep{2000A&A...360..952H,1996A&A...313..497F} commonly used in MESA \citep{2011ApJS..192....3P}, enables the reproduction of asteroseismic constraints and leads to abundance profiles at boundaries, which are comparable to those of 3D simulations \citep{2017ApJ...836L..19A}. This exponentially-decaying mixing leads to larger convective cores like penetrative overshoot or entrainment but at a rate/extent which may not be correct. We may thus have to include the growth rate of convective regions (e.g. via entrainment) and the process that determines the shape of their boundaries (e.g. shear mixing) as separate processes in 1D codes in order to reproduce all the properties of convective boundaries found in 3D hydrodynamic simulations. Further steps along the 321D road include replacing the mixing-length theory \citep{1958ZA.....46..108B} with a theory able to predict convective properties such as velocities in the entrained region \citep[see discussion in][]{2015ApJ...809...30A}.

\section*{Acknowledgements}
\cdre{The authors would first like to thank the referee for their continued questioning, concerns and suggestions, resulting in a greatly improved manuscript.} The authors acknowledge support from EU-FP7-ERC-2012-St Grant 306901. RH acknowledges support from the World Premier International Research Centre Initiative (WPI Initiative), MEXT, Japan. 
This article is based upon work from the “ChETEC” COST Action (CA16117), supported by COST (European Cooperation in Science and Technology). AC acknowledges partial support from NASA Grant NNX17AG24G. This research used resources of the
National Energy Research Scientific Computing Center (NERSC), which is
supported by the Office of Science of the U.S.  Department of Energy
under Contract No.  DE-AC02-05CH11231.
This work used the Extreme Science and Engineering Discovery Environment (XSEDE), which is supported by National Science Foundation grant number OCI-1053575. CM and WDA acknowledge support from NSF grant 1107445 at the 
University of Arizona. The authors acknowledge the Texas Advanced Computing Center (TACC) at The University of Texas at Austin (http://www.tacc.utexas.edu) for providing HPC resources that have contributed 
to the research results reported within this paper. This work used the DiRAC Data Centric system at 
Durham University, operated by the Institute for Computational Cosmology on behalf of the STFC DiRAC HPC Facility (www.dirac.ac.uk). This equipment was funded by BIS National E-infrastructure capital grant 
ST/K00042X/1, STFC capital grants ST/H008519/1 and ST/K00087X/1, STFC DiRAC Operations grant ST/K003267/1 and Durham University. DiRAC is part of the National E-Infrastructure. \corr{The authors acknowledge PRACE for awarding access to the resource MareNostrum 4 based in Spain at Barcelona Supercomputing Center. The support of David Vicente and Janko Strassburg from Barcelona Supercomputing Center, Spain, to the technical work is gratefully acknowledged. CG, RH, and CM acknowledge ISSI, Bern, for its support in organising a collaboration meeting which sparked many fruitful discussions. This research has made use of NASA's Astrophysics Data System Bibliographic Services.}

\newpage
\begin{appendix}
\section{Computational tools}\label{comp_tools}
Initial conditions for the 3D hydrodynamic simulations are provided by a stellar evolution model of a 15\,M$_\odot$ star evolved up to the point of carbon shell burning. This model was calculated using the Geneva stellar evolution code (\textsc{genec}). The 3D hydrodynamic simulations presented here were calculated using the Prometheus MPI (\textsc{prompi}; where MPI is an acronym for message passing interface) code. Both codes are described below.\\ 
\noindent The 1D stellar model structure and composition were mapped onto a finer grid for use in the \textsc{prompi} code. These finely zoned stellar model data can be found in full in machine-readable format as supplementary online material. Several zones from these initial conditions are also shown in Table \ref{1d_model_IC}.

\begin{table*}
\centering
\begin{tabulary}{\textwidth}{l || c c c c c c c c}
\hline\hline\\
\textsf{i} & \textsf{r [cm]} & \textsf{s [erg\,K$^{-1}$]} & \textsf{X$_{^{12}\rm C}$} & \textsf{$\bar{A}$} & \textsf{$\bar{Z}$} & \textsf{$\rho$ [g\,cm$^{-3}$]} & \textsf{T [K]} & \textsf{P [bar]}\\\\
\hline \hline\\
0001 & 4.161$\times10^8$ & 1.649$\times10^8$ & 6.099$\times10^{-3}$ & 18.17 & 9.072 & 1.737$\times10^6$ & 8.642$\times10^8$ & 9.594$\times10^{22}$ \\\\
0002 & 4.165$\times10^8$ & 1.650$\times10^8$ & 6.099$\times10^{-3}$ & 18.17 & 9.072 & 1.733$\times10^6$ & 8.643$\times10^8$ &           9.570$\times10^{22}$ \\\\
0003 & 4.169$\times10^8$ & 1.651$\times10^8$ & 6.099$\times10^{-3}$ & 18.17 & 9.072 & 1.729$\times10^6$ & 8.644$\times10^8$ &           9.546$\times10^{22}$ \\\\
.\;.\;.\\\\
1249 & 8.987$\times10^8$ & 2.979$\times10^8$ & 5.056$\times10^{-2}$ & 17.63 & 8.807 & 1.870$\times10^5$ & 9.043$\times10^8$ &           9.805$\times10^{21}$\\\\
1250 & 8.990$\times10^8$ & 3.046$\times10^8$ & 6.852$\times10^{-2}$ & 17.42 & 8.699 & 1.826$\times10^5$ & 9.201$\times10^8$ &           9.868$\times10^{21}$\\\\
1251 & 8.994$\times10^8$ & 3.123$\times10^8$ & 8.877$\times10^{-2}$ & 17.18 & 8.578 & 1.773$\times10^5$ & 9.361$\times10^8$ &           9.895$\times10^{21}$\\\\
.\;.\;.\\\\
4998 & 2.348$\times10^9$ & 3.947$\times10^8$ & 3.571$\times10^{-1}$ & 14.38 & 7.183 & 8.171$\times10^3$ & 3.472$\times10^8$ &           1.717$\times10^{20}$\\\\
4999 & 2.349$\times10^9$ & 3.948$\times10^8$ & 3.571$\times10^{-1}$ & 14.38 & 7.183 & 8.163$\times10^3$ & 3.471$\times10^8$ &           1.715$\times10^{20}$\\\\
5000 & 2.349$\times10^9$ & 3.948$\times10^8$ & 3.571$\times10^{-1}$ & 14.38 & 7.183 & 8.155$\times10^3$ & 3.471$\times10^8$ &           1.713$\times10^{20}$\\\\
\hline\hline
\end{tabulary}
\caption{Truncated table of the re-zoned stellar model initial conditions used in the \textsc{prompi} simulations. From left to right the following variables are listed: zone number, radius, entropy, $^{12}$C abundance, mean atomic mass, mean atomic number, density, temperature and pressure. Values are shown for zones within the bottom of the computational domain (top rows), lower convective boundary (middle rows) and the top of the computational domain (bottom rows). The full table of values is available in machine-readable form as supplementary online material.}
\label{1d_model_IC}
\end{table*}

\subsection{The Geneva stellar evolution code}\label{genec}

\indent \textsc{genec} \citep{2008Ap&SS.316...43E, 2012A&A...537A.146E} solves the stellar evolution equations (see e.g. Eqs. 10.1 - 10.5 of \citealt{2013sse..book.....K}) using a finite difference, time implicit method within a Lagrangian framework. 

The chemical composition is homogeneously mixed in convective regions up to oxygen burning. The structure equations are de-coupled from the abundance equations; changes in abundances due to nuclear burning and diffusive mixing are calculated separately. 

Reaction rates are calculated for a nuclear reaction network of 23 isotopes. Energy losses due to the production and loss of neutrinos is included. A perfect gas equation of state including radiation and partial degeneracy is used. Opacities are interpolated from tables provided by the OPAL group \citep{1994ApJ...437..879A,1996ApJ...456..902R}. Mass loss estimates are a function of metallicity, and calculated according to the prescriptions by \citet{2001A&A...369..574V} and \citet{1988A&AS...72..259D}. Energy transport due to convection is assumed to be adiabatic for deep internal convection, and for convective envelopes, energy transport is modelled using the mixing length theory with $\alpha_{ml} = 1.6$ \citep{1992A&AS...96..269S}. The extent of convectively unstable regions, is determined by the Schwarzschild criterion, along with penetrative convection of up to 10\% of the local pressure scale height (for core hydrogen and helium burning phases), where the pressure scale height is $H_p=-d r/d \,\textrm{ln}\, p=p/g\rho$.

\subsection{The Prometheus MPI code}\label{prompi}


\textsc{prompi} \citep{2007ApJ...667..448M} is a finite-volume, time explicit, Eulerian code which uses the piecewise-parabolic method of \citet{1984JCoPh..54..174C}. 

\subsubsection{Piecewise-Parabolic Method}\label{ppm}

\cdre{The piecewise-parabolic method (PPM) is a higher order extension of Godunov's method \citep{zbMATH03273813, GODUNOV19621187}, which utilises a forward Euler method for time integration\footnote{\cdre{Note that, PPM is a special method in the sense that it has a higher effective resolution than other differencing schemes, e.g. DNS \citep[see e.g. ][for a discussion]{2000JCoPh.158..225S}}}. In multiple dimensions PPM is second order accurate in space and time \citep{1990JCoPh..87..171C}. PPM provides a Kolmogorov description of the turbulent cascade down to the dissipation level at the sub-grid level. It was developed with shock capturing in mind, and relates the change in turbulent kinetic energy and traversal time across a shock to the specific entropy production rate \citep{1962JFM....13...82K,2007iles.book.....G}, or} 
\begin{equation}
\frac{\partial E_K}{\partial t} = -\frac{\partial s}{\partial t} = \frac{5}{4}\frac{\Delta v^3}{\Delta r}.
\end{equation}
\cdre{This is the definition for the energy dissipation rate along a turbulent cascade (in a statistically steady state, see also Appendix \ref{reeff}). Without the use of an explicitly defined viscosity PPM matches motions on the turbulent cascade to those at the grid scale. In the ILES paradigm then, the physical scale at which dissipation occurs is replaced by the grid scale. The conservation laws are enforced to machine accuracy, and so the turbulent cascade is well represented (see Fig. \ref{ke_spectra_eps} and fig. 8 of \citealt{2017MNRAS.471..279C}), providing that the inertial range is sufficiently sampled on the grid.}

\subsubsection{\textsc{prompi} design}\label{prompi-design}

\noindent \textsc{prompi} is derived from the legacy astrophysics code \textsc{prometheus} \citep{1989nuas.conf..100F}. \textsc{prompi} utilises domain decomposition in order to be parallelised over multiple processors. Communication between computational nodes is handled by MPI. 

\corrdre{The Euler equations for fluid motion (inviscid approximation) are solved within the ILES paradigm \citep[e.g.][]{2007iles.book.....G}},

\begin{align}
\frac{\partial\rho}{\partial t} + \boldsymbol{\nabla}\cdot(\rho \,\boldsymbol{v}) &= 0;\\
\rho\,\frac{\partial \boldsymbol{v}}{\partial t} + \rho\,\boldsymbol{v}\cdot\boldsymbol{\nabla}\boldsymbol{v} &= -\boldsymbol{\nabla}p +\rho\,\mathbf{g};\\
\rho\,\frac{\partial E_t}{\partial t} + \rho\,\boldsymbol{v}\cdot\boldsymbol{\nabla}E_t + \boldsymbol{\nabla}\cdot(p\,\boldsymbol{v})&=\rho\,\boldsymbol{v}\cdot\mathbf{g}+\rho(\epsilon_{nuc}+\epsilon_\nu);\\
\rho\,\frac{\partial X_i}{\partial t} + \rho\,\boldsymbol{v}\cdot\boldsymbol{\nabla}X_i &= R_i,
\end{align}

\noindent \corrdre{where $p$ is the pressure, $\mathbf{g}$ the gravitational 
acceleration, $E_t$ the total energy, $X_i$ the mass fraction of nuclear species $i$ and $R_i$ the rate of change of 
nuclear species $i$}.

\subsubsection{Model set-up}

\corrdre{\citet{1941DoSSR..30..301K} showed that the rate of TKE dissipation at all scales does not depend on the details of the dissipative process. This implies that it may be unnecessary to resolve the dissipation sub-range of the cascade, 
meaning that the sub-grid dissipation becomes the effective Kolmogorov length, through its implied viscosity. 
In a turbulent flow the momentum diffusion (viscosity) is negligible in comparison to the advection of KE at all scales except near the Kolmogorov scale and below it. In ILES, the minimum amount of dissipation required in order to maintain monotonicity and energy conservation is adopted.}

\corrdre{The implicit sub-grid scale model used in ILES is the leading order term in the truncation error, which is a result of discretising the Euler equations \cdre{through the use of PPM, and is therefore of second order accuracy\footnote{\cdre{Unlike other LES methods, ILES does not require the use of an explicit sub-grid model.}}.} The qualitative effects of dissipation at the grid scale are implemented in ILES. Energy conservation is built into the ILES scheme; KE that cascades down to the sub-grid scales from the resolved scales is damped as the velocity fluctuations are dissipated. The internal energy of the fluid is suitably increased in such a way that it mimics viscosity at the Kolmogorov scale, which would dissipate the structures at this scale into heat.}

\cdre{The specific form of this numerical diffusion is the same as the simplest case described in the appendix of \citet{1984JCoPh..54..174C}. That is the interpolation functions are flattened based on the steepness of the pressure jump across a zone and an extra explicit diffusive flux is also added. In multiple dimensions, this method is valid providing that the same flattening is applied to the derivatives in each direction \citep{1990JCoPh..87..171C}.}

A Cartesian geometry is used, and the boundary conditions are reflective in the vertical direction and periodic in the horizontal directions. Velocities are also damped near the domain boundary in the vertical direction using a sinusoidal function. Explicit time-stepping is used to ensure that all structural and acoustic changes within cells are temporally resolved. 

\corrdre{The time-step, $\Delta t$, is restricted such that it is in accordance with the CFL condition \citep{1928MatAn.100...32C}, specifically}

\begin{equation}
\Delta t = C_{\rm max} \frac{\Delta x}{c_s},
\end{equation}

\corrdre{where $\Delta x$ is the cell width, $c_s$ is the local sound speed and $C_{\rm max}$ is the Courant factor set to be 0.8.}

The Helmholtz EOS \citep{1999ApJS..125..277T,2000ApJS..126..501T} is used to describe the thermodynamic state of the plasma. Convective heat transport is initiated in the model through small, random and equal perturbations in temperature and density. Energy generation from nuclear fusion is parameterised as a function of composition, density and temperature, \corrdre{using a slightly modified version of the parameterisation given by \citet{1986nce..conf.....A,2009pfer.book.....M}:}

\begin{equation}
\epsilon_{^{12}\rm C} \sim 4.8\times10^{18}\;Y_{12}^2\; \rho\; \lambda_{12,12},
\end{equation}\label{enuc}

\noindent \corrdre{where $Y_{12} = X_{^{12}\rm C}/12$, $X_{^{12}\rm C}$ is the mass fraction of $^{12}$C, $\lambda_{12,12} = 5.2\times10^{-11}\, T_9\hspace{0.05mm}^{30}$ and 
$T_9=T/10^9$.} The loss of energy due to escaping neutrinos is parameterised using the analytical formula provided by \citet{1967ApJ...150..979B},

\begin{align*} 
\epsilon_\nu = & \frac{1.590\times10^{14}\lambda^8}{2(1+25.22\lambda)}\\ & + \frac{21.6\rho^2\lambda^2}{\rho+8.6\times10^5}(1+2.215\times10^{-6}\xi^2)e^{-4.5855\times10^{-3}\xi}\\ & + 4.772\times10^2\lambda^2\rho\, e^{-2.5817\times10^{-5}\rho^{2/3}\lambda^{-1}},  
\end{align*} 

\corrdre{where $\lambda=k_BT/mc^2$ and $\xi=\rho^{1/3}\lambda^{-1}$.}

\section{Effective Reynolds number derivation}\label{reeff}

\corrdre{A useful dimensionless number for determining the extent of turbulence in a simulation is the effective Reynolds number. This is the discrete analogue of the Reynolds number, $\textrm{Re}=v\, \ell/\nu$, where $v$ and $\ell$ are the velocity and length scales and $\nu$ is the viscosity. The effective Reynolds number is defined using the following arguments. The rate of energy dissipation at a length scale $\lambda$ is $\epsilon\sim v^3/\lambda$ \citep{1941DoSSR..30..301K}. One can then approximate the rates of energy dissipation at the extreme scales of the simulation, i.e. at the integral scale and the grid scale. These energy dissipation rates are}

\begin{equation}\label{eps_scales}
\epsilon_{\ell}=\frac{v_{\rm rms}^3}{\ell}\;\;\;\textrm{and}\;\;\; \epsilon_{\Delta x}=\frac{\Delta v^3}{\Delta\, x},\;\;\;\textrm{respectively},
\end{equation}\label{epsilon_scale}

\corrdre{where $\Delta v$ is the flow velocity 
across a grid cell; this velocity can also be used to define an effective numerical viscosity at the grid scale}

\begin{equation}\label{nu_eff}
\nu_{\rm eff}=\Delta v \Delta x.
\end{equation}

\corrdre{For a turbulent system within a statistically steady state, \citet{1962JFM....13...82K} showed that the rate of energy dissipation is equal at all scales; applying this equality to Eq. \ref{eps_scales} yields (with the use of Eq. \ref{nu_eff})}

\begin{equation}
\nu_{\rm eff}=v_{\rm rms}\ell \left(\frac{\Delta x}{\ell}\right)^{4/3}.
\end{equation}

\corrdre{Assuming the integral scale is the size of the convective region, $\ell_{cz}$, then the effective Reynolds number can be expressed as (to within a factor of 2)}

\begin{equation}\label{re_eff}
\textrm{Re}_{\rm eff}=\left(\frac{\ell}{\Delta x}\right)^{4/3} \sim N_x^{\,4/3},
\end{equation}

\corrdre{where $N_x$ is the number of grid points in the vertical direction. In these simulations this is a slight over-estimate as in the vertical direction only half of the grid points represent the convective region. Therefore, for these models which have a vertical resolution of $N_x=512$ zones (256 in the convective zone), the effective Reynolds number is, Re$_{\rm eff}\sim 1625 - 4000$, which is comfortably within the turbulent regime.}

\section{Specific kinetic energy spectra}\label{app_spectra}

\begin{figure}
\includegraphics[width=0.5\textwidth]{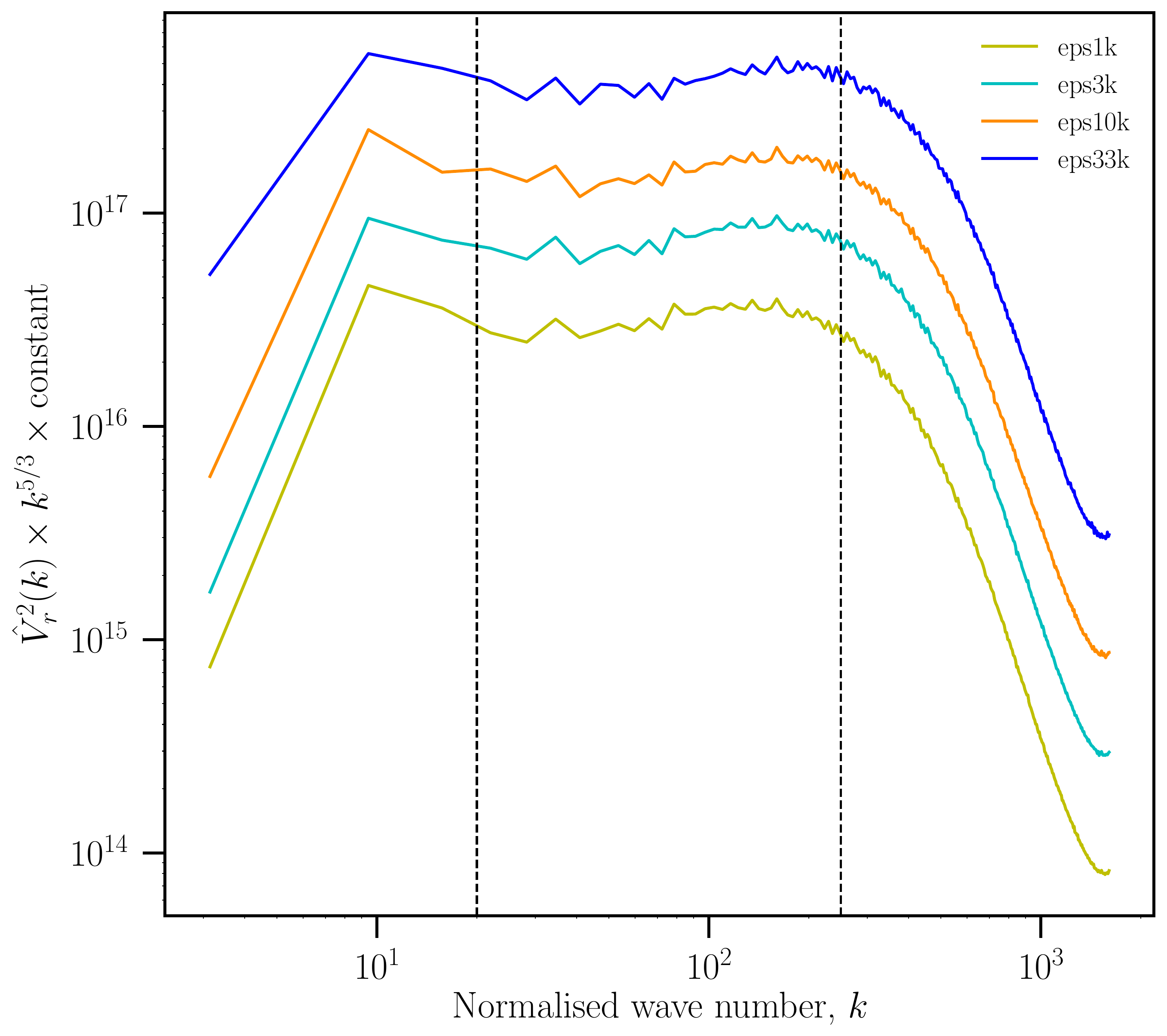} 
\caption[Kinetic energy spectra of the \textsf{eps1k} - \textsf{eps33k} models]{Specific kinetic energy spectrum for the \textsf{eps1k}, \textsf{eps3k}, \textsf{eps10k} and \textsf{eps33k} models in the luminosity study. Spectra were obtained from the squared 2D Fourier transform (in the centre of the convective region) of the vertical velocity. The vertical axis consists of the Fourier transform, scaled by a constant. The horizontal axis is the normalised wave-number, $k=\sqrt{k_y^2+k_z^2}$. The vertical dashed lines represent the approximate wavenumber range of three different regions, the integral range (left), inertial range (centre) and dissipation range (right).}
\label{ke_spectra_eps}
\end{figure} 

The properties of the inertial sub-range of scales within the turbulent cascade are explored by comparing velocity spectra for each model at different luminosities. This is achieved by performing a horizontal 2D fast Fourier transform\footnote{Using the Python package \textsc{numpy.fft.fft2}.} (FFT) of the vertical velocity at a constant height, within the centre of the convection zone. The results of this transform are presented in Fig. \ref{ke_spectra_eps}, where the transform squared, $\hat{V}(k)^2$, 
is plotted as a function of the wave-number. These spectra are time-averaged over several convective turnover times, where the convective turnover time is given by, $\tau_c=2\ell_{cz}/v_{\rm rms}$, with $\ell_{cz}$ being the height of the convective region (see Table\,\ref{eps_tab} for values of $\tau_c$). The 1D profile is obtained by binning the 2D transform within the $k_y-k_z$ plane, where $k_y$ and $k_z$ are the wave-numbers in the y and z directions, respectively ($k_y, k_z=0,2\pi,4\pi,...,2\pi(N/2),$ where $N$ is the number of grid points in one dimension, i.e. the resolution). A scaling of $(1/N_yN_z)$ is applied, where $N_y$ and $N_z$ are the two horizontal resolutions. The spectra have also been normalised by $k^{5/3}$ in order to depict the regions of the spectra that follow Kolmogorov's $k^{-5/3}$ power law \citep{1941DoSSR..30..301K} as roughly horizontal. This region is roughly located between the two vertical dashed lines at approximate wave numbers of $k=20$ and $k=250$. This wave-number range represents the scales of the models that are no longer affected by the initial or boundary conditions, driving force or dissipation. We also attribute the leftmost region as the integral range and the rightmost region as the dissipation range. 
The magnitude of the specific kinetic energy (velocity squared) increases with driving luminosity, as would be expected.\\
\indent \corrdre{Velocity power spectra at various resolutions up to $1024^3$ are calculated by \citet{2000JCoPh.158..225S} for both PPM simulations solving the Euler equations and up to $512^3$ solving the Navier-Stokes (NS) equations with explicit viscosity terms. They showed that by comparing the kinetic energy time dependence that the PPM solutions required around 4 times less resolution than the NS solutions to model the same system.}\\

\section{Reynolds-averaged Navier Stokes framework}\label{ssec:rans}

It is common, when studying turbulent flows, to average the governing equations both spatially, to obtain a mean turbulent state over two dimensions, and temporally, to smooth out the stochastic nature of turbulence and provide a statistical average. When taking a Reynolds average of the Euler equations, a new ‘mean evolution’
of the fluid flow can be represented by averaging 
the horizontal components into a
radial one. 

Reynolds decomposition by construction separates the mean flow component from the fluctuating component.  One can then construct physically relevant terms, which represent competing processes for a given
conservation law \citep[Ch.5 of][]{chassaing_2002}. 
The mean field is calculated by averaging over the horizontal
plane, i.e. for a quantity $\omega$

\begin{equation}
\left<\omega\right>=\frac{1}{\Delta A}\int_{\Delta A}\omega \,\textrm{d}A,
\end{equation}

where $\textrm{d}A = \textrm{d}y\,\textrm{d}z$ and $\Delta A = \Delta y\,\Delta z$ is the area of the computational domain
perpendicular to the radial axis. The original quantity can then be expressed as the sum of the mean and fluctuating components

\begin{equation}
\omega=\overline{\left<\omega\right>}+\omega',
\end{equation}

where the Reynolds average of the fluctuation is by definition zero, i.e. $\overline{\left<\omega'\right>}=0$. The overbar notation denotes a temporal average over the quasi-steady state. This provides a better, statistically more meaningful representation of the flow. It also smooths out many of the instantaneous
fluctuations provided the number of convective turnovers simulated is large enough ($\gtrsim4$). This temporal averaging is defined as:

\begin{equation}
\overline{\omega}=\frac{1}{\Delta t}\int_{t_1}^{t_2}\omega(t) dt
\end{equation}

for an averaging window $\Delta t=t_2-t_1$.

\begin{figure*}
\includegraphics[width=0.75\textwidth]{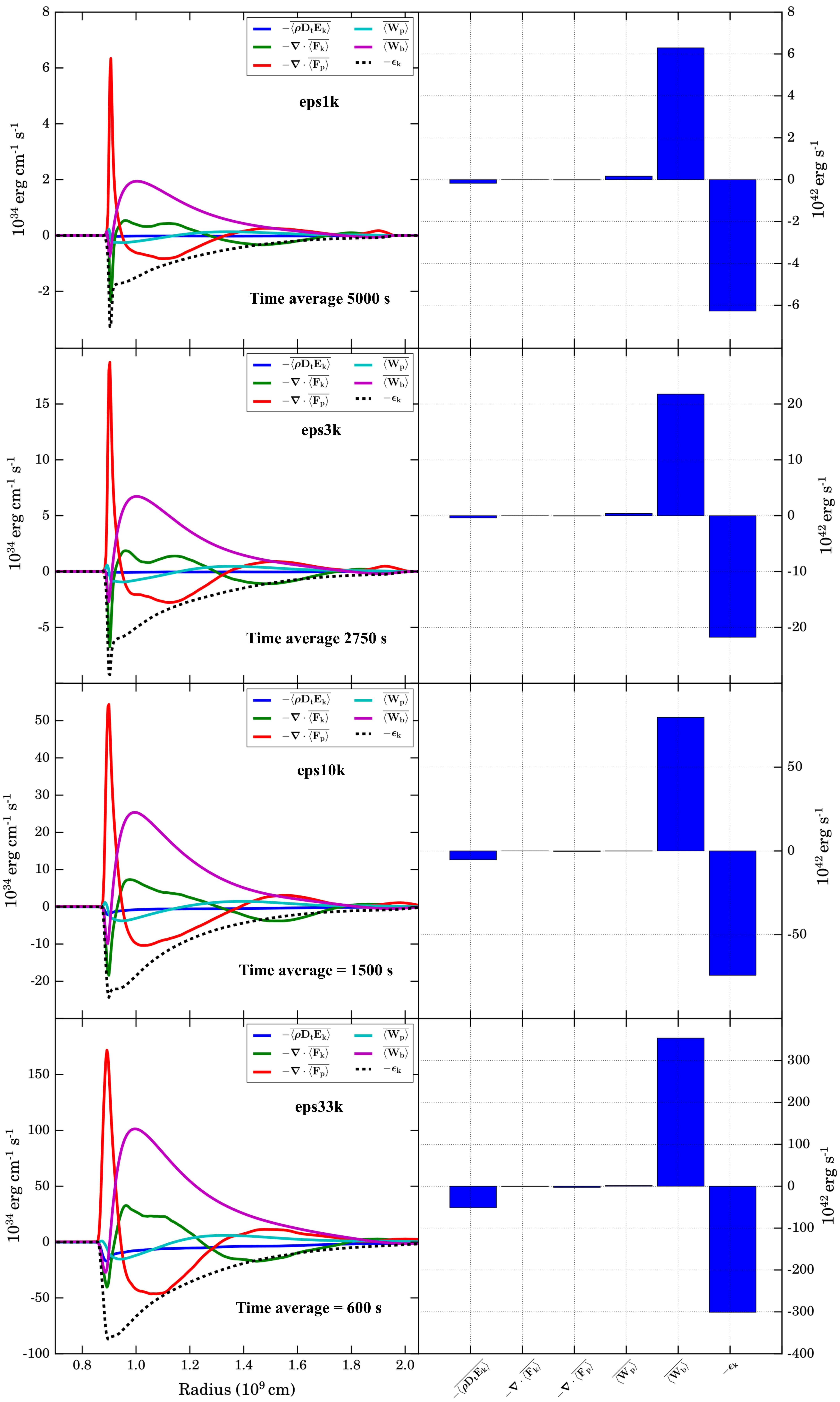}
\caption[TKE budget of the \textsf{eps1k} and \textsf{eps3k} models]{\textsf{Left}: Decomposed terms of the mean kinetic energy equation (Eq. \ref{mmfka}), which have been 
horizontally averaged, normalised by the domain surface area and time averaged over the quasi-steady state period. Time averaging 
windows are 5,000$\,$s, 2,750$\,$s, 1,500\,s and 600\,s for the \textsf{eps1k} (top panels), \textsf{eps3k} (upper middle), \textsf{eps10k} (lower middle) and \textsf{eps33k} (bottom) models, 
respectively. \textsf{Right}: Bar charts representing the radial integration of the profiles shown in the 
left panel. \protect{This plot is analogous to fig. 8 of \citet{2013ApJ...769....1V}.}}
\label{mfa_ke_1k_3k}
\end{figure*} 

\begin{figure*}
\centering
\includegraphics[width=0.49\textwidth]{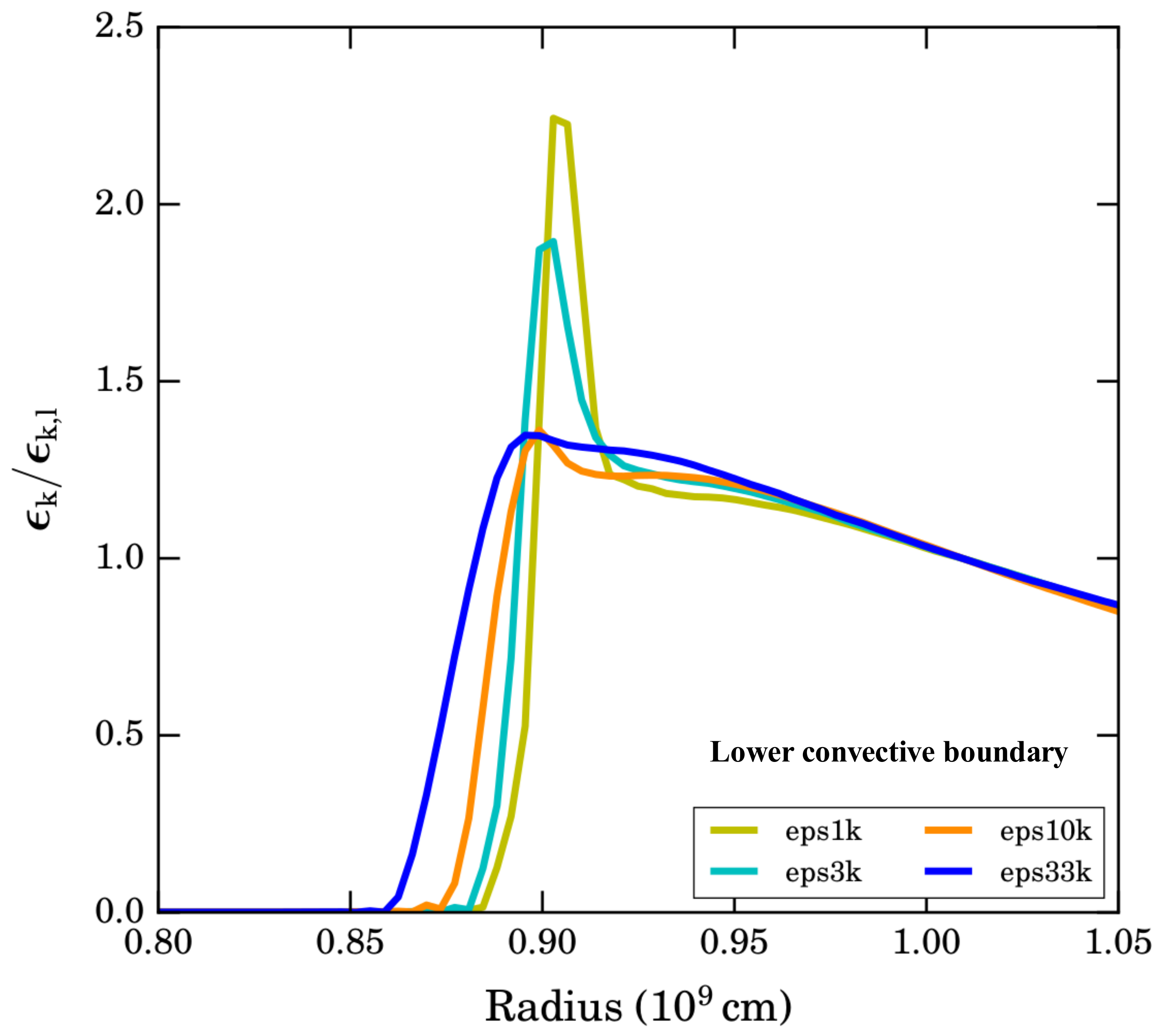}
\includegraphics[width=0.49\textwidth]{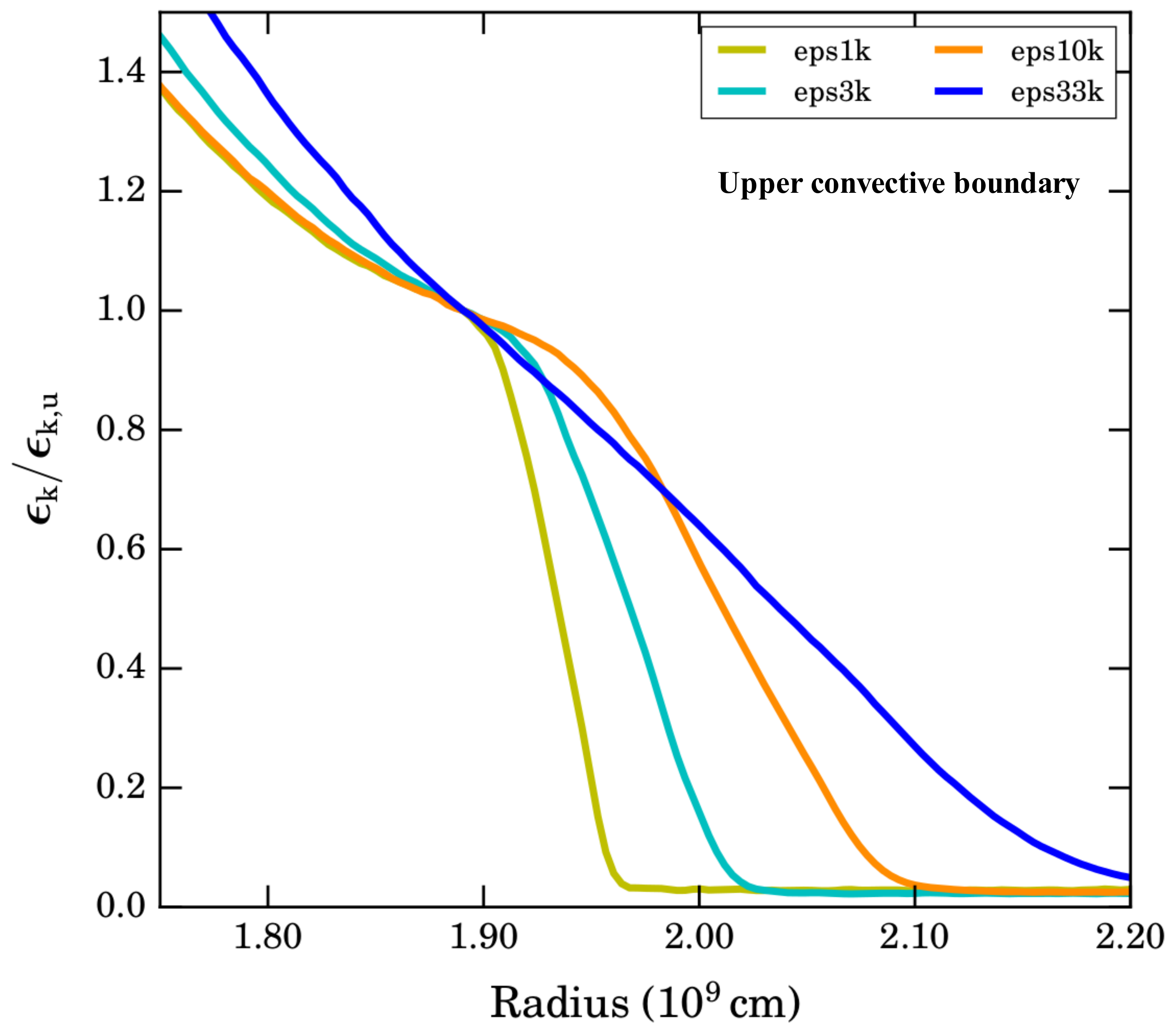}
\caption[TKE residual dissipation profiles of the \textsf{eps1k} - \textsf{eps33k} models]{Turbulent kinetic energy residual dissipation curves of the lower (left) and upper (right) convective boundary regions
for the: \textsf{eps1k} (yellow); \textsf{eps3k} (cyan); \textsf{eps10k} (orange); and \textsf{eps33k} (blue) models. The numerical dissipation for each boundary has been normalised 
by a value at a common position within the convective region near to the boundary ($\sim1.0\times10^9\,$cm and $\sim1.9\times10^9\,$cm). Therefore, any sensitivity to the spatial resolution should be revealed by sharp peaks in the dissipation profiles, of which there are none at the upper boundary, but all models except the \textsf{eps33k} model possess sharp peaks at the lower boundary.}
\label{eps_eku}
\end{figure*} 

\subsection{Turbulent Kinetic Energy Equation}\label{app-tke}
The Eulerian equation of turbulent kinetic energy can be written as (eq. A12 of \citealt{2007ApJ...667..448M}):

\begin{equation}
\partial_t\left(\rho E_k\right)+ \boldsymbol{\nabla}\cdot\left(\rho E_k\boldsymbol{v}\right)=-\boldsymbol{v}\cdot\boldsymbol{\nabla}p+\rho\,\boldsymbol{v}\cdot\boldsymbol{g},
\end{equation}

where $v$ is the velocity and $E_k=\frac{1}{2}(v\cdot v)$ is the specific kinetic energy.

Applying horizontal and temporal averaging to the above equation yields the mean turbulent kinetic energy equation, which can be written as,

\begin{align}
\overline{\left<\rho \mathbf{D_t} E_k\right>}=&-\boldsymbol{\nabla}\cdot\overline{\left<\mathbf{F_p}+\mathbf{F_k}\right>}+\overline{\left<\mathbf{W_p}\right>}+\overline{\left<\mathbf{W_b}\right>}-\epsilon_k;
\end{align}\label{mmfka}

where $\mathbf{D_t} =\partial_t + \boldsymbol{v} \cdot \boldsymbol{\nabla}$ is the material derivative;

$\mathbf{F_p}=p'\boldsymbol{v'}$ is the turbulent pressure flux;

$\mathbf{F_k}=\rho E_k\boldsymbol{v'}$ is the TKE flux;

$\mathbf{W_p}=p'\boldsymbol{\nabla}\cdot\boldsymbol{v'}$ is the pressure dilatation;

$\mathbf{W_b}=\rho'\mathbf{g}\cdot\boldsymbol{v'}$ is the work due to buoyancy; and

$\epsilon_k$ is the implied numerical dissipation of KE.

\subsection{Effects of driving luminosity on the turbulent kinetic energy budget}\label{tke_eps}

The turbulent kinetic energy (TKE) budget of the models in this driving luminosity study are interpreted within the Reynolds-averaged Navier-Stokes (RANS) framework. All of the terms in Eq. \ref{mmfka} were calculated for the \textsf{eps1k} - \textsf{eps33k} models  and are shown in Fig. \ref{mfa_ke_1k_3k}. The radial profiles of each term are presented in the left panels of this figure, with the inferred viscous dissipation shown by a dashed line. \corr{This numerical dissipation is calculated from the residual TKE ($\epsilon_k$ in Eq. \ref{mmfka}). For a well resolved turbulent system, it should be a smooth profile.} Each profile is time averaged over multiple convective turnover times and normalised by the surface area of the domain. The right panels show bar charts of the radial integration of each term shown in the left panel. It can be seen that all profiles increase in magnitude as the driving luminosity is increased, simply because more nuclear energy is being put into the system. 

The main driving term for the TKE is the buoyancy work, $\mathbf{W_b}$, which for the \textsf{eps1k, eps3k} and \textsf{eps10k} models is balanced by the numerical dissipation at the grid scale, $\epsilon_k$. Due to such a balance, the shell is in a statistically steady state, as shown by the very small values of the material time derivative of the TKE, $\rho \mathbf{D_t} E_k$. For the \textsf{eps33k} model, the driving term is so large that while most of the TKE is dissipated, some of the energy affects the shell dynamically. Towards the end of the simulation, it is no longer in a statistically steady state and the shell itself is eventually completely disrupted. 

The peak in the residual TKE (dashed curve \corr{in Fig. \ref{mfa_ke_1k_3k}}) at the bottom of the convective shell decreases in amplitude with increased driving luminosity, and the profile also broadens with increased driving luminosity. This is mainly due to the broadening of the boundary itself as the boundary is more easily overwhelmed by turbulent motions with an increased driving luminosity. For the most energetic model, \textsf{eps33k}, the lower boundary is broad enough that the spatial resolution is sufficient to model this boundary physically and without any adverse numerical effects due to poor resolution. 

To investigate this in more detail, the residual TKE is compared between these four models for the upper and lower boundary in Fig. \ref{eps_eku}. In this figure the numerical dissipation is normalised by a value at a common position within the convective region close to the boundary. This normalisation value is numerically converged at these spatial resolutions (512$^3$) and so poor spatial resolution at the boundary is revealed by peaks in the dissipation curves. 
For the upper boundary (right panel in Fig. \ref{eps_eku}), the absence of any peaks suggests that the adopted resolution is adequate to resolve this boundary of these four models. With the exception of the \textsf{eps33k} model, all of the models steadily decrease in relative dissipation towards the boundary, followed by a steeper slope beyond the boundary, and eventually plateauing to a very small value within the stable upper region. For the \textsf{eps33k} model, however, the TKE is so great that the reduction in dissipation across the boundary region occurs at an almost constant slope. A possible explanation for such behaviour is that the TKE of this model is so great that it destroys the boundary layer altogether (see Fig. \ref{vmag_eps33k}).\\
\indent The lower boundary is much narrower than the upper boundary (see e.g. Table \ref{tabbw_eps}). This is also apparent in Fig. \ref{eps_eku} from the appearance of spurious peaks in the relative dissipation curves in all models except \textsf{eps33k}. The increase in driving luminosity broadens the lower boundary, resulting in an increase in the effective resolution for the higher energy models. This broadening reduces the amplitude of the dissipation peaks. In the \textsf{eps10k} model the dissipation peak is small and in the \textsf{eps33k} it is almost non-existent. As expected, a lower spatial resolution is needed to resolve a broader boundary. 
\corrdre{We see that an increase in the luminosity at a fixed resolution explored in this study has the same positive effect on resolving the boundary as an increase in resolution at a fixed luminosity (as done in C17).}

\section{Equilibrium entrainment regime}\label{equil_entrain}
\corrdre{Turbulent mixing is considered to occur within the equilibrium entrainment regime if the time scale for the boundary migration due to entrainment, $\tau_b$, is comparable to or larger than the convective turnover time scale, $\tau_c$ \citep{2004JAtS...61..281F,2014AMS...1935.JGJG}. These time scales are defined as}

\begin{equation}\label{tau_c}
\tau_c=\frac{2\ell_{cz}}{v_{\rm rms}},
\end{equation}

\begin{equation}\label{tau_c}
\tau_b=\frac{\delta r}{|v_e|},
\end{equation}

\corrdre{where $\delta r$ is the boundary width (see Table \ref{tabbw_eps}) and $v_e$ is the entrainment velocity (see Fig. \ref{boundary_pos_ind}). The time scales for each boundary and the ratio $\tau_b$/$\tau_c$ are 
given 
for the \textsf{eps1k}, \textsf{eps3k}, \textsf{eps10k} and \textsf{eps33k} models in Table \ref{tab_ee}. This ratio is moderately larger than one for all of these boundaries suggesting that turbulent entrainment occurs within the equilibrium entrainment regime.}

\begin{table}
\centering
\begin{tabulary}{0.5\textwidth}{l || c c c c}
\hline\hline\\
 & \textsf{eps1k} & \textsf{eps3k} & \textsf{eps10k} & \textsf{eps33k}\\\\
\hline \hline\\
\textsf{$\tau_{b,u} (\rm s)$}  & 7066  & 3377  & 1346 & 662\\\\
\textsf{$\tau_{b,l} (\rm s)$}  & 15843  & 7649  & 3391 & 975\\\\
\textsf{$\tau_c (\rm s)$}  &  465  &  316 & 219 & 153\\\\
\textsf{$\tau_{b,u}/\tau_c$}  & 15.2  & 10.7 & 6.1 & 4.3\\\\
\textsf{$\tau_{b,l}/\tau_c$}  & 34.1  & 24.2 & 15.5 & 6.4\\\\
\hline\hline
\end{tabulary}
\caption{{Convective turnover times, boundary migration time scales and the ratio of these two time scales for each boundary of the \textsf{eps1k}, \textsf{eps3k}, \textsf{eps10k}, \textsf{eps33k} models.}}
\label{tab_ee}
\end{table}

\section{Bulk Richardson number}\label{ssec:rib}

The bulk Richardson number is a useful diagnostic for the susceptibility of a boundary region to the entrainment of material through turbulent motions (more broadly categorised as convective boundary mixing). We refer to this susceptibility as the stiffness of the boundary, i.e. a stiffer boundary is less susceptible to convective boundary mixing. The bulk Richardson number is defined as the ratio of the specific stabilisation potential (analogous to the work done against convective motions by the boundary) to the specific TKE within the convective region. It is written as

\begin{equation}\label{rib}
\textrm{Ri}_B = \frac{\Delta B \ell}{v_{\rm rms}^2},
\end{equation}

where $\Delta B$ is the buoyancy jump across the boundary. Based on the results of \citet{2007ApJ...667..448M}, we take the integral length scale, $\ell$, to be half of the local pressure scale height at the boundary. The buoyancy jump is estimated by integrating the 
square of the buoyancy (Brunt-V\"{a}is\"{a}l\"{a}) frequency, $N$, over a suitable distance ($\Delta r$) either side of the boundary centre, 
$r_c$,

\begin{equation}\label{buoy}
\Delta B=\int\limits_{r_c-\Delta r}^{r_c+\Delta r}N^2dr.
\end{equation}

The integration distance $\Delta r$ is not well defined theoretically but it should be large enough to capture the dynamics of the boundary region and the distance over which fluid 
elements are decelerated. In these simulations we take the integration distance, $\Delta r$, to be a quarter of the local pressure scale height at the boundary.

The buoyancy frequency is the frequency with which a perturbed fluid element will oscillate at if it is surrounded by a stably stratified medium. This frequency is imaginary for a convectively unstable fluid element and is defined as:

\begin{equation}\label{n2}
N^2 = \frac{g}{\rho}\left(\frac{\partial \rho}{\partial s}\frac{ds}{dr} + \frac{\partial \rho}{\partial \bar{A}}\frac{d\bar{A}}{dr} + \frac{\partial \rho}{\partial \bar{Z}}\frac{d\bar{Z}}{dr}\right).
\end{equation}

Bulk Richardson numbers less than 10 are associated with relatively soft convective boundaries, whereas bulk Richardson numbers greater than 100 are associated with relatively stiff convective boundaries.

\end{appendix}

\newpage

\bibliography{references}

\bsp

\label{lastpage}

\end{document}